\documentclass[11pt]{article}
\pdfoutput=1
\usepackage{jcapmod}

\usepackage{shorthand}
\usepackage{mathtools}
\usepackage{booktabs}
\usepackage[english]{babel}
\usepackage{amsmath,amssymb,amsbsy,amstext, amsthm, simplewick, amsfonts}
\usepackage{graphicx}
\usepackage[small]{caption}
\usepackage{siunitx}
\usepackage{upgreek}
\usepackage{framed}
\usepackage{wrapfig}
\usepackage{multirow}
\usepackage{bbm}
\usepackage[svgnames,dvipsnames,x11names]{xcolor}
\usepackage{nicematrix}

\usepackage{array}

\newcolumntype{C}[1]{>{\centering\arraybackslash}m{#1}}

\usepackage{selinput}

\usepackage{bm}
\usepackage{float}
\usepackage{geometry}
\usepackage{yfonts}
\usepackage{subcaption}
\usepackage{sidecap}
\usepackage{longtable}
\usepackage{anyfontsize}
\usepackage{dsfont}
\usepackage{tikz}
\usetikzlibrary{decorations.markings, arrows.meta}
\tikzset{
  midarrow/.style={
    postaction={
      decorate,
      decoration={
        markings,
        mark=at position 0.5 with {\arrow{Stealth}}
      }
    }
  }
}
\usepackage{relsize}
\usepackage{tcolorbox}

\usepackage{xparse}

\usepackage{slashed}
\usepackage{simpler-wick}

\NewDocumentCommand{\colornucleus}{omme{_^}}{%
  \begingroup\colorlet{currcolor}{.}%
  \IfValueTF{#1}
   {\textcolor[#1]{#2}}
   {\textcolor{#2}}
    {%
     #3
     \IfValueT{#4}{_{\textcolor{currcolor}{#4}}}
     \IfValueT{#5}{^{\textcolor{currcolor}{#5}}}
    }%
  \endgroup
}

\newcolumntype{L}[1]{>{\raggedright\let\newline\\\arraybackslash\hspace{0pt}}m{#1}}
\newcolumntype{C}[1]{>{\centering\let\newline\\\arraybackslash\hspace{0pt}}m{#1}}
\newcolumntype{R}[1]{>{\raggedleft\let\newline\\\arraybackslash\hspace{0pt}}m{#1}}

\usepackage[framemethod=default]{mdframed}
\newmdenv[skipabove=7pt,
skipbelow=7pt,
rightline=false,
leftline=false,
topline=false,
bottomline=false,
backgroundcolor=gray!10,
linecolor=gray,
innerleftmargin=5pt,
innerrightmargin=5pt,
innertopmargin=5pt,
innerbottommargin=5pt,
leftmargin=0cm,
rightmargin=0cm,
linewidth=4pt]{eBox}
\newmdenv[skipabove=7pt,
skipbelow=7pt,
rightline=false,
leftline=false,
topline=false,
bottomline=false,
backgroundcolor=gray!10,
linecolor=gray,
innerleftmargin=5pt,
innerrightmargin=5pt,
innertopmargin=-5pt,
innerbottommargin=5pt,
leftmargin=0cm,
rightmargin=0cm,
linewidth=4pt]{eBox2}

\definecolor{Red}{RGB}{214, 39, 40}
\definecolor{Blue}{RGB} {31, 119, 180}
\definecolor{Orange}{RGB}{255, 153, 51}
\definecolor{Purple}{RGB}{178, 102, 255}
\definecolor{Green}{RGB}{44, 160, 44}

\definecolor{vio}{RGB}{19, 130, 164}
\definecolor{vioo}{RGB}{89, 2, 155}
\newcommand{\Comment}[1]{{}}
\definecolor{darkblue}{rgb}{0.15,0.35,0.55}
\definecolor{reddish}{rgb}{0.65, 0.2, 0.2}
\definecolor{darkgreen}{RGB}{50,150,0}
\definecolor{greyish}{rgb}{.90,.90,.90}
\definecolor{greyish2}{rgb}{.96,.96,.96}
\definecolor{greyish3}{rgb}{.37,.37,.37}
\definecolor{darkblue2}{rgb}{0.3,0.4,0.9}
\definecolor{Blue3}{RGB}{31, 119, 180}
\usepackage[linktocpage=true]{hyperref}
\hypersetup{
colorlinks=true,
citecolor=darkblue,
linkcolor=reddish,
urlcolor=darkblue,
pdfauthor={},
pdftitle={},
pdfsubject={}
}

\usepackage{colortbl}
\definecolor{lightgreen}{cmyk}{0.2, 0, 0.2, 0.2}
\definecolor{lightgray2}{cmyk}{0.1,0.1,0,0.1}
\definecolor{Red2}{RGB}{214, 39, 40}
\definecolor{Blue2}{RGB} {31, 119, 180}
\definecolor{Orange2}{RGB}{255, 127, 14}
\definecolor{Green2}{RGB}{44, 160, 44}

\makeatletter
\newlength{\apb@width}
\newcommand{\autoparbox}[2][c]{\settowidth{\apb@width}{#2}\parbox[#1]{\apb@width}{#2}}

\makeatother


\def\hs{\hskip 1pt}

\def\beq{\begin{equation}}
\def\eeq{\end{equation}}
\def\be{\begin{equation}}
\def\ee{\end{equation}}

\def\CC{\hat C}

\newcommand{\ud}{{\rm d}}

\newcommand{\dif}{\mathrm{d}}
\let\CC\relax
\newcommand{\CC}{\mathbb{C}}
\newcommand{\RR}{\mathbb{R}}
\newcommand{\OGr}{\mathrm{OGr}}

\DeclarePairedDelimiter\floor{\lfloor}{\rfloor}

\allowdisplaybreaks[1]
\theoremstyle{definition}
\newtheorem{definition}{Definition}[]
\theoremstyle{remark}

\newcommand{\ogr}{{\rm OGr}}
\newcommand{\gr}{{\rm Gr}}
\newcommand{\PP}{\mathbb{P}}

\newcommand{\Pf}{{\rm Pf}}

\begin{document}

\newgeometry{top=2cm, bottom=2cm, left=2cm, right=2cm}

\begin{titlepage}
\setcounter{page}{1} \baselineskip=15.5pt 
\thispagestyle{empty}

\begin{center}
{\fontsize{21}{18} \bf Positive Geometry of Yang--Mills Correlators}
\end{center}

\vskip 20pt
\begin{center}
\noindent
{\fontsize{14}{18}\selectfont 
Mattia Arundine\hs$^{1,2}$, Veronica Calvo Cortes\hs$^{3}$, Joris Koefler\hs$^{3}$ and Facundo Rost\hs$^{4}$}
\end{center}

\begin{center}
\vskip8pt
\textit{$^1$ Institute of Physics, University of Amsterdam, Amsterdam, 1098 XH, The Netherlands}
\vskip8pt
\textit{$^2$  Leung Center for Cosmology and Particle Astrophysics,
Taipei 10617, Taiwan}
\vskip8pt
\textit{$^3$  Max Planck Institute for Mathematics in the Sciences, Leipzig 04103, Germany}
\vskip 8pt
\textit{$^4$ Scuola Normale Superiore and INFN, Piazza dei Cavalieri 7, 56126, Pisa, Italy}
\end{center}

\vspace{0.4cm}
\begin{center}{\bf Abstract}
\end{center}
\noindent
We develop a positive-geometric formulation of tree-level Yang--Mills correlators in de Sitter space at three and four points through their helicity-stripped representatives on the cosmological Grassmannian.
In its Pfaffian (or spinor) embedding, physical singularities become natural geometric boundaries.
At three points, the Yang--Mills correlator is the canonical form of the non-negative orthant in the Grassmannian.
At four points, the Mandelstam divisors partition the Pfaffian-positive domain of the Grassmannian into four positive geometries. Requiring factorization into three-point forms, together with the correct flat-space limit, uniquely selects an oriented union of two of these regions, whose canonical form reproduces the reduced color-ordered Yang--Mills correlator. The full color-ordered correlator, on the other hand, arises from a uniquely fixed signed linear combination of homology cycles. Thus, the broader homological formulation of positive geometry is essential for capturing the complete four-point result. Our construction provides a concrete starting point for a geometric description of higher-point cosmological correlators.

\end{titlepage}
\restoregeometry

\newpage
\setcounter{tocdepth}{3}
\setcounter{page}{2}

\linespread{1.2}
\tableofcontents
\linespread{1.1}

\newpage
\section{Introduction}
Cosmological correlators are fundamental observables of the early Universe: they encode the primordial fluctuations that set the initial conditions for its subsequent evolution and ultimately seed late-time structure~\cite{Baumann:2022jpr}. Conventionally, they are computed perturbatively by evolving quantum fields in de Sitter space. Even at low orders, these calculations involve intricate time integrals, while the resulting expressions are often remarkably simple. This tension suggests that the conventional dynamical description may not be the most natural starting point for understanding these observables.

\vskip 4pt
One may instead ask whether cosmological correlators can be determined directly from their kinematic and dynamical constraints, rather than through the usual perturbative calculations. In such a formulation, the correlator would arise as the answer to an intrinsically defined problem, without reference to time evolution, while the familiar principles of bulk physics would emerge from its consistency conditions. Besides providing a potentially more efficient route to their computation, such a perspective could offer a fundamentally new way of formulating and understanding quantum field theory in de Sitter space, as well as the primordial initial conditions of our Universe~\cite{Arkani-Hamed:2018kmz}.

\vskip 4pt
Indeed, symmetries, singularities, factorization and the flat-space limit impose stringent constraints on cosmological correlators.
Their singularities can occur only on particular loci, and their behavior near these loci is fixed by lower-point observables or by scattering amplitudes~\cite{Arkani-Hamed:2018kmz, Baumann:2020dch, Baumann:2021fxj, Goodhew:2020hob, Goodhew:2021oqg}. On a factorization locus, for example, the leading singularity of an $n$-point correlator is determined by lower-point correlators. On the flat-space locus, instead, its leading singularity reproduces a scattering amplitude. Additional properties, such as color ordering, further restrict which singularities may occur~\cite{Berends:1987me, Mangano:1990by, Dixon:1996wi, DelDuca:1999rs, Brandhuber:2022qbk}.

\vskip 4pt
These requirements are not independent. Different kinematic loci intersect, so the corresponding limits must be mutually compatible. An iterated residue can be computed by approaching an intersection through different sequences of boundaries, and the resulting expressions must agree.
The singularity structure of a correlator therefore constitutes a recursive system: higher-point functions reduce to lower-point functions on factorization loci, which in turn satisfy analogous conditions on their own singularities.

\vskip 4pt
This suggests regarding a cosmological correlator as the solution to a geometric problem.
The physical singularities specify the boundaries of the geometric space, whereas factorization and the flat-space limit determine the behavior there.
Positive geometry provides a natural framework for precisely this problem, as has been shown for scattering amplitudes and cosmological correlators alike~\cite{Arkani-Hamed:2017tmz, brown2025positivegeometriescanonicalforms, Arkani-Hamed:2009hub, Arkani-Hamed:2009ljj, Arkani-Hamed:2012zlh, Arkani-Hamed:2013jha, Huang:2013owa, Huang:2014xza, Elvang:2014fja, He:2023rou, Damgaard:2019ztj, Arkani-Hamed:2017mur, Arkani-Hamed:2018ign, Arkani-Hamed:2017fdk, Arkani-Hamed:2024jbp, Benincasa:2024leu}.
Given a real oriented region $R$ in an ambient space $X$, its canonical form $\omega_R$ has logarithmic singularities along the boundaries of $R$. On every codimension-$1$ boundary $B\subset\partial R$, it obeys the recursive relation
\be
    \mathop{\rm Res}_{B=0}\omega_R=\omega_B\,,
\ee
up to an orientation convention. Iterating this relation associates intersections of boundaries with iterated residues of the canonical form. This leads to a direct dictionary:
\begin{align*}
\text{kinematic singularities}
\ &\longleftrightarrow\
\text{geometric boundaries}, \\
\text{physical limits}
\ &\longleftrightarrow\
\text{residues of canonical forms}.
\end{align*}
Factorization is then interpreted as the statement that the canonical form on a boundary is assembled from lower-point canonical forms, while the flat-space limit identifies a distinguished boundary stratum with scattering amplitude data.

\vskip 4pt
The positive geometry program goes beyond providing a geometric rewriting of a known answer. Ultimately, the goal is that the geometric constraints fully determine the function that we are interested in computing.
Once the allowed boundaries and the canonical forms on them are specified, the recursive residue conditions may determine the form in the interior. The geometry can therefore organize the analytic constraints on a correlator and, when uniqueness holds, reconstruct the correlator from its kinematic limits. Conversely, an incompatibility among the proposed boundaries or their residues provides a geometric obstruction to the corresponding analytic structure.

\vskip 4pt
The main challenge is to identify an ambient space in which this dictionary can be implemented. Such a space should make the symmetries of the correlator manifest, represent its physical singularities as algebraic loci, and allow the corresponding boundaries to be combined consistently. The cosmological Grassmannian introduced in~\cite{Arundine:2026fbr} (see also \cite{Arundine:2026myr, Huang:2026tsh, Bala:2026hdm, Bala:2026bdx, Bala:2026lvw}) provides a natural candidate: it recasts cosmological correlators as differential forms on a geometric kinematic space, where their symmetries are trivialized and their singularity structure can be studied independently of many of the complications of their conventional momentum-space expressions.

\vskip 4pt
In this paper, we take this geometric viewpoint as our starting point. We identify positive regions in the cosmological Grassmannian whose boundaries represent the allowed kinematic limits, and require their canonical forms to satisfy factorization and the flat-space limit. No theory-specific correlator is supplied to this construction. Since a canonical form is fixed by its boundary data, these requirements leave essentially no freedom: at three and four points, the resulting forms turn out to be precisely the (tree-level) correlators of Yang--Mills theory. Yang--Mills thus emerges only at the end of the procedure, as the most natural theory compatible with the positive-geometric realization of the prescribed kinematic limits.\footnote{We observe that tree-level correlators with external gluons are unable to distinguish between Yang--Mills theory and its supersymmetric extensions.}

\subsection*{Strategy and Summary}
We now sketch the main results of this work and the steps taken to derive them, while reserving the details and the explicit computations for the main text.

\vskip 4pt
We consider $n$-point wavefunction coefficients $\psi_n$, which encode essential information about correlation functions on the future (conformal) boundary of four-dimensional (Anti-)de Sitter space (A)dS$_4$. While our presentation will focus on de Sitter space, the extension to AdS is straightforward.

\vskip 4pt
In de Sitter space, the isometries of the bulk act on the three-dimensional future boundary as conformal transformations. Furthermore, wavefunction coefficients of gauge fields satisfy the same conservation equations as currents on the boundary. Conformal symmetry and current conservation of wavefunction coefficients can therefore be leveraged to represent the latter in the following integral form~\cite{Arundine:2026fbr}:
\be
    \label{eq:introint}
    \psi_n(\Lambda) = \int \dif C \: \delta(C \cdot Q \cdot C^T) \, \delta(C\cdot\Lambda) \, A_n(C) \,,
\ee
where $\Lambda$ is a $2n\times 2$ matrix of spinors that encodes the momenta of the external fields and $C$ is an $n \times 2n$ real matrix that satisfies the orthogonality constraint $C \cdot Q \cdot C^T = 0$, with
\be
    Q = \begin{pmatrix}
        0 & 1_{n\times n} \\
        1_{n\times n} & 0
    \end{pmatrix}.
\ee
Geometrically, the rows of $C$ identify a collection of $n$ vectors in $\mathbb{R}^{2n}$, and the matrix is associated with the $n$-dimensional plane that these span. This defines the \textit{orthogonal Grassmannian} $\ogr(n,2n)$ as the space of $n$-dimensional $Q$-null planes in $\mathbb{R}^{2n}$.
The matrices $C$ are defined up to $\mathrm{GL}(n)$ transformations $C \mapsto R \cdot C$, since such transformations merely correspond to a change of basis. Because of the quadratic constraint $C \cdot Q \cdot C^T = 0$, which admits two different ``square roots'' as solutions, this space features two disconnected isomorphic branches. For any choice of external helicities in Yang--Mills theory, the correlator is non-zero on only one of the two branches, so we can treat them independently. For an even number of negative helicities, a possible representation of the associated branch is $C = (1_{n \times n}, \, C_n)$, with $C_n$ a skew-symmetric $n \times n$ matrix whose entries are $(C_n)_{ij} = -c_{ij}$.

\vskip 4pt
The integral representation \eqref{eq:introint} allows us to focus our attention on the Grassmannian correlator $A_n(C)$, which takes a much simpler form than $\psi_n(\Lambda)$. Invariance under $\rm{GL}(n)$ transformations forces this function to depend only on the determinants $(I_1 \cdots I_n)$ of the $n \times n$ submatrices of $C$, with $I_i \in \{ \bar 1, \bar 2, \dots , \bar n,1,2,\dots,n \}$ labeling the columns of $C$. Covariance under little group transformations of the spinor matrix $\Lambda$, on the other hand, implies a homogeneity constraint on $A_n(C)$ that depends on the external helicities of the fields. For $n=3$, the answer is fixed by the external helicities alone. For example, the Yang--Mills three-point wavefunction coefficient in the chart $C = (1_{n \times n}, \, C_n)$ is~\cite{Baumann:2024ttn, Arundine:2026fbr}
\be
    \label{eq:A3intro}
    A_3 = \frac{1}{c_{12} \hs c_{13} \hs c_{23}} \,,
\ee
where we have stripped the helicity-dependent numerator. For $n=4$, on the other hand, extra dynamical information is needed. First, we define the Mandelstam variables~\cite{Arundine:2026fbr}
\be
    \label{eq:Mandeldef}
    S = (\bar 1 \bar 2 1 2) \,, \quad T = (\bar 1 \bar 4 1 4) \,, \quad U = (\bar 1 \bar 3 1 3) \,.
\ee
A generic \emph{tree-level} four-point function can have poles when each Mandelstam vanishes. Theories like Yang--Mills, on the other hand, allow the assignment of a ``color'' to each external field. This operation preserves only a cyclic relabeling of the external fields, so that the associated $A_4(C)$ has no singularity at $U = 0$. On the singularities at $S = 0$ and $T = 0$, the function $A_4(C)$ factorizes into products of three-point functions $A_3$. It must also hold that $A_4(C)$ reduces to a scattering amplitude on the locus $(\bar1\bar2\bar3\bar4) = (1234) = 0$. More concretely, for the Yang--Mills four-point function, one has
\be
    \mathop{\rm Res}_{(1234)=(\bar1\bar2\bar3\bar4)=0} A_4(C) \propto \mathcal{M}_4(\mathcal{C})\,,
\ee
where $\mathcal{C}$ is an element of $\gr(2,4)$, i.e.~the space of $2$-planes in $\mathbb{R}^4$, and $\mathcal{M}_4$ is the Yang--Mills scattering amplitude. The color-ordered Yang--Mills correlator is~\cite{Arundine:2026fbr}
\be
    \label{eq:A4intro2}
    A_4 = \frac{1}{S \hs T} \bigg( \frac{3}{S+T+U} + \frac{1}{S+T-U} - \frac{1}{-S+T+U} - \frac{1}{S-T+U} \bigg) \,,
\ee
where again we have stripped the polarization structure. We can further define the following ``reduced'' Yang--Mills correlator\footnote{Both Grassmannian correlators are mapped to different discontinuities of the Yang--Mills four-point function, as discussed in~\cite{Arundine:2026fbr}.}
\be
    \label{eq:A4intro}
    \hat A_4 = \frac{1}{2} \Big( A_4(U) + A_4(-U) \Big) = \frac{4 \hs (S+T)}{(S+T+U)(S+T-U) S \hs T}\,.
\ee
Both functions are compatible with the flat-space limit and the required factorization property.

\vskip 4pt
In this paper, we will show that these results have a deeper geometric origin. We will first define ``positive'' regions in the orthogonal Grassmannian, with boundaries given by the kinematic singularities of the wavefunction coefficient.
We will prove that these regions are positive geometries and that their canonical forms indeed reproduce the Yang--Mills correlators determined on physical grounds in~\cite{Baumann:2024ttn, Arundine:2026fbr}.

\vskip 4pt
The task of characterizing the positive geometries underlying Yang--Mills correlators is best addressed via an equivalent representation of (one branch of) $\ogr(n,2n)$ as a subvariety in $\mathbb{P}^{m}$, with $m = 2^{n-1}-1$.
As a first example, we observe that the right branch of $\ogr(3,6)$ is isomorphic to $\mathbb{P}^3$. The isomorphism is obtained via the map $(1_{3 \times 3}, \, C_3) \mapsto (1: c_{12} : c_{13} : c_{23})$, which uplifts to the homogeneous Pfaffian coordinates $(p : p_{12} : p_{13} : p_{23})$. Similarly, for $n=4$, we can map $(1_{4 \times 4}, \, C_4) \mapsto (1: c_{12} : \cdots : c_{34}:c_{1234})$, with $c_{1234} = c_{12}c_{34} - c_{13}c_{24} + c_{14}c_{23}$, which uplifts to the coordinates $(p: p_{12}: \cdots:p_{34}:p_{1234})$. These are not independent, but lie on the hypersurface
\be
    p \hs p_{1234} - p_{12} \hs p_{34} + p_{13} \hs p_{24} - p_{14} \hs p_{23} = 0\,.
\ee
We interpret each $p_{ij}$ as a sub-Pfaffian of $-C_n$, i.e.~as the square root of the determinant of the submatrix with rows and columns in $\{ i,j \}$.
Analogously, $p_{1234}$ is the Pfaffian of $-C_4$. 
As we explain in Appendix~\ref{app: Spinors}, the projective Pfaffian coordinates can be seen as components of a (projective) spinor that satisfies the so-called purity constraints~\cite{Hughston:1988nz,Berkovits:2004bw}.

\vskip 4pt
For $n=3$, the positive region $R$ is simply the positive orthant of $\mathbb{P}^3$, where all the coordinates are non-negative, $p, \hs p_{ij} \geq 0$. This is an elementary positive geometry, whose canonical form is
\be
    \label{eq:w3intro}
    \omega_3 = \frac{1}{p_{12} \hs p_{13} \hs p_{23}} \, \dif^3p_{ij} \,.
\ee
We see that the canonical function in \eqref{eq:w3intro} precisely matches \eqref{eq:A3intro} in the projective chart $p = 1$, where $p_{ij} \mapsto c_{ij}$.

\begin{figure}
    \centering
    \includegraphics[width=0.42\linewidth]{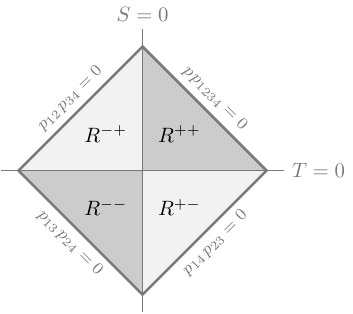}
    \caption{Sketch of the positive orthant of (the right branch of) $\OGr(4,8)$ and the four regions $R^{\pm \pm}$ arising from the $S = 0$ and $T = 0$ hypersurfaces. Each region is bounded by $S = 0, \: T = 0$, and the vanishing of two Pfaffian coordinates. As explained in the main text, this two-dimensional picture is obtained via a projection from the original six dimensions. Here, the two Pfaffian boundaries of each $R^{\pm\pm}$ collapse to a single line. This picture differs from the rectangle found in~\cite{Bala:2026lvw} by a choice of coordinates.}
    \label{fig:introfig}
\end{figure}

\vskip 4pt
The case of $n=4$ requires more care. Again, we begin by restricting to the subset with non-negative coordinates, $p, \hs p_{ij}, \hs p_{1234} \geq 0$. We note that the Mandelstams defined in \eqref{eq:Mandeldef} can be written as
\be
\begin{aligned}
    S & = p_{12} \hs p_{34}- p \hs p_{1234} = p_{13} \hs p_{24} - p_{14} \hs p_{23} \,, \\[4pt]
    T & = p_{14} \hs p_{23}- p \hs p_{1234} = p_{13} \hs p_{24} - p_{12} \hs p_{34}\,.
\end{aligned}
\ee
The loci $S = 0$ and $T = 0$ introduce additional boundaries in the kinematic space. These hypersurfaces split the positive orthant into four isomorphic regions $R^{\pm \pm}$, as shown in Figure~\ref{fig:introfig}. These regions differ by the signs of $S$ and $T$, and are individually positive geometries. The uniqueness of the canonical form of each of the four regions (and of any of their unions) is guaranteed by verifying that the hypotheses of the Brown-Dupont theorem~\cite{brown2025positivegeometriescanonicalforms} are satisfied.

\vskip 4pt
To reproduce the physical correlator $A_4$, the $6$-form $\omega_4$ must factorize into a wedge product of two $3$-forms $\omega_3$ on the two hypersurfaces $S=0$ and $T = 0$, modulo a $\rm{GL}(1)$ redundancy. Demanding further that the double residue on the codimension-$2$ locus $p = p_{1234} = 0$ of $\omega_4$ returns the $4$-form of a scattering amplitude on the non-negative Grassmannian $\gr(2,4)_{\geq 0}$ forces us to examine the region $R^{++}$. This region has both $p=0$ and $p_{1234} = 0$ as boundaries, but on its own does not satisfy the necessary factorization property on either $S = 0$ or $T = 0$. The correct flat-space limit and the correct factorization on \textit{both} $S = 0$ and $T = 0$ then lead to the union of $R^{++}$ with $R^{--}$. This is a positive geometry, whose canonical form is
\be
    \omega_4(R^{++} \cup R^{--}) = \frac{-(S+T)}{p \hs p_{1234} \hs p_{13} \hs p_{24} \hs S \hs T} \, \dif^6 p_{ij} \,.
\ee
This is the \textit{only} union of a region with $R^{++}$ that features both $S=0$ and $T=0$ as boundaries, and its canonical function equals the ``reduced'' Yang--Mills correlator \eqref{eq:A4intro} due to the equalities $S+T+U = -2p \hs p_{1234}$ and $S+T-U = 2p_{13} \hs p_{24}$. Notably, this result is forced on us by the physical requirements we have imposed, which leave no room for the geometric construction of other correlators.

\vskip 4pt
The geometric nature of the reduced correlator and its uniqueness imply that a different strategy must be adopted to obtain the full color-ordered correlator \eqref{eq:A4intro2}. Our solution is to relax the rigidity of the problem, and consider a different linear combination of the canonical forms of the fundamental regions $R^{\pm\pm}$. This time, we allow this linear combination to contain arbitrary rational coefficients, rather than restricting to the binary values $\{ 0, \pm 1\}$ that follow from simply joining (oriented) regions together. Imposing also the correct factorization property, we find the unique combination
\begin{align}
    \label{eq:introHodgeform}
    \Omega_4 & = \frac{1}{2}\Big[ \omega_4(R^{++}) - \omega_4(R^{--}) \Big] + \frac{1}{2}\Big[ \omega_4(R^{++})+\omega_4(R^{+-}) \Big] + \frac{1}{2}\Big[ \omega_4(R^{++})+\omega_4(R^{-+}) \Big] \nonumber\\[4pt]
    & = \frac{3}{2}\omega_4(R^{++}) - \frac{1}{2}\omega_4(R^{--}) + \frac{1}{2}\omega_4(R^{+-}) + \frac{1}{2}\omega_4(R^{-+}) \,,
\end{align}
which corresponds to the oriented sum of the three possible reduced correlators, one for each Mandelstam. While not a positive geometry in the original sense, as there is no natural semi-algebraic region corresponding to $\Omega_4$, this sum can naturally be interpreted as a positive geometry using the mixed Hodge theory framework. Up to the factor of $1/2$, furthermore, each term in the first line of \eqref{eq:introHodgeform} also identifies a positive geometry in the original sense.

\vskip 4pt
It is intriguing that the answers to these geometric questions naturally lead us to rediscover the Grassmannian representatives of the Yang--Mills correlators found in~\cite{Baumann:2024ttn, Arundine:2026fbr}, and make us wonder what type of structures await at higher points.

\subsection*{Outline}
The paper is organized as follows. In Section~\ref{sec:cosmoGrass}, we review the representation of wavefunction coefficients as integrals over the cosmological Grassmannian and isolate the Grassmannian correlator that contains their dynamical information. We discuss its relation to scattering amplitudes through the flat-space limit and introduce the Pfaffian coordinates used throughout the paper.

\vskip 4pt
In Section~\ref{sec:positive_geometries}, we review positive geometry through a series of examples. We present both its original formulation, in which canonical forms are characterized by their residues on boundaries, and the broader formulation in terms of relative (co)homology. The latter allows canonical forms to be associated with cycles that need not correspond to individual semi-algebraic regions.

\vskip 4pt
In Section~\ref{sec:YM_from_geometry}, we apply these ideas to regions of the cosmological Grassmannian whose boundaries encode the relevant kinematic limits. We prove that these regions define positive geometries and determine their canonical forms. Requiring the appropriate factorization and flat-space behavior singles out the three- and four-point Yang--Mills correlators. The complete color-ordered four-point result requires the homological formulation of positive geometries. Finally, we recast the four-point construction as a planar hyperplane arrangement. We establish the isomorphism between the two descriptions and use the arrangement to determine the dimension of the space of logarithmic top-forms with singularities along the kinematic boundaries.

\vskip 4pt
We conclude in Section~\ref{sec:conclusions}.

\vskip 4pt
In Appendix~\ref{app: Spinors}, we explain how to interpret the Pfaffian coordinates used in this work as the components of projective pure spinors.
To this end, we first introduce the Clifford algebra of $\mathbb R^{n,n}$ and realize it as the algebra of $n$ fermionic oscillators.
We then describe explicitly the one-to-one correspondence between the orthogonal Grassmannian and the space of projective pure spinors. The appendix is self-contained and can be read independently of the main text.

\section{Cosmological Grassmannian}
\label{sec:cosmoGrass}
In this section, we review the cosmological Grassmannian as a kinematic space for wavefunction coefficients of massless spinning fields in four-dimensional de Sitter space, whose boundary kinematics are those of conserved currents in three dimensions. We begin with their integral representation over the orthogonal Grassmannian and isolate the Grassmannian correlator, which contains the dynamical information. We then discuss the constraints imposed on this function by $\mathrm{GL}(n)$ covariance and little group scaling, as well as factorization on its physical singularities and the reduction to scattering amplitudes in the flat-space limit. We illustrate these structures using Yang--Mills correlators as our main example. Finally, we introduce the Pfaffian, or spinor, embedding of the orthogonal Grassmannian, which furnishes the natural coordinate system in which we will study positive regions and their boundaries in the remainder of the paper.

\subsection{Wavefunction Coefficients}
Particles in the bulk of de Sitter space are described as fluctuations of quantum fields with spin. Interactions among them are imprinted in the statistical properties of these fields on the future boundary of the spacetime, which are encoded in their $n$-point correlation functions. These, in turn, can be determined from auxiliary objects, the $n$-point wavefunction coefficients. The wavefunction coefficients of massless spinning fields are special, as they satisfy the same kinematic properties as correlation functions of conserved currents in a three-dimensional conformal field theory.

\vskip 4pt
Leveraging their kinematic properties, it was shown in~\cite{Arundine:2026fbr} that (discontinuities of) wavefunction coefficients of massless fields admit the following integral representation:
\be
    \label{eq:masterint}
    \psi_n(\Lambda) = \int \frac{\dif^{n \times 2n} C}{\mathrm{GL}(n)} \, \delta(C \cdot Q \cdot C^T) \, \delta(C \cdot \Lambda) \, A_n(C) \,.
\ee
We will now explain each element appearing in this formula. First, $\Lambda$ is the following $2n \times 2$ \textit{real} matrix:
\be
    \Lambda = \begin{pmatrix}
        \lambda_1^1 & \lambda_1^2 \\
        \vdots & \vdots \\
        \lambda_n^1 & \lambda_n^2 \\
        \bar \lambda_1^1 & \bar\lambda_1^2 \\
        \vdots & \vdots \\
        \bar \lambda_n^1 & \bar \lambda_n^2
    \end{pmatrix} \,,
\ee
where $\lambda_i^\alpha$ and $\bar \lambda_i^\alpha$, with $\alpha = 1,2$, are spinors that describe the three-dimensional momentum $\vec{k}_i$ of the $i$-th field.\footnote{For the details of this map, we refer to~\cite{Arundine:2026fbr}.} The momenta are invariant under the \textit{little group transformations} $\lambda_i \mapsto \rho \lambda_i,  \: \bar \lambda_i \mapsto \rho^{-1} \bar \lambda_i$, with $\rho \in \mathbb{R}^*$. 
The integral is performed over $n\times 2n$ real matrices $C$ of full rank that satisfy the quadratic constraint $C \cdot Q \cdot C^T = 0$, with
\be\label{equ:Q}
    Q = \begin{pmatrix}
        0 & 1_{n \times n} \\
        1_{n \times n} & 0
    \end{pmatrix}.
\ee
The representation \eqref{eq:masterint} can therefore be interpreted as an integral over the \emph{orthogonal Grassmannian} $\ogr(n,2n)$, the space of $Q$-null $n$-planes in $\RR^{2n}$. A full-rank matrix $C$ represents the plane spanned by its rows, and the quotient by $\mathrm{GL}(n)$ identifies matrices related by $C\mapsto R\cdot C$, with $R\in\mathrm{GL}(n)$, since left multiplication changes only the choice of basis for this plane. We label the columns of $C$ by $\bar 1,\ldots,\bar n,1,\ldots,n$. The quadratic constraint $C \cdot Q \cdot C^T = 0$ implies that $\OGr(n,2n)$ is an algebraic variety with two connected components, called the ``left'' and ``right'' branches.
Let $V_{C}$ be the $n$-plane spanned by the rows of $C$, and $C_0\equiv (1_{n\times n},0_{n\times n})$ a reference matrix. We then define the branches as
\be
\begin{aligned}
    \OGr_R(n,2n) &\equiv\{ V_C \subset \RR^{2n}  \colon \dim (V_{C_0} \cap V_C) \equiv n  &\mod 2\} \,,\\[4pt]
    \OGr_L(n,2n) &\equiv\{ V_C\subset \RR^{2n}  \colon \dim (V_{C_0} \cap V_C) \equiv n+1 \hspace{-10pt} & \mod 2\} \,.
\end{aligned}
\ee
The presence of the delta function $\delta(C \cdot \Lambda)$ ensures that $\psi_n(\Lambda)$ is a distribution supported on the locus where three-momentum is conserved:
\be
    \psi_n(\Lambda) \propto \delta(\Lambda^T \cdot Q \cdot \Lambda) \propto \delta \bigg(\sum_{i=1}^n \vec{k}_i\bigg) \,.
\ee
Finally, the \textit{Grassmannian correlator} $A_n(C)$ in \eqref{eq:masterint} encodes the same physical information as the wavefunction coefficient $\psi_n(\Lambda)$. We will discuss its properties in the next subsection.

\vskip 4pt
Note that all Yang--Mills correlators with an even number of negative helicity gluons localize in the right branch. In this connected component, we can gauge-fix the $\mathrm{GL}(n)$ redundancy by considering the following form of the Grassmannian matrix:
\be
    \label{eq:rightchart}
    C = \begin{pmatrix}
        1_{n \times n}\,, & C_n
    \end{pmatrix},
\ee
where $C_n$ is a skew-symmetric $n \times n$ matrix, with $n(n-1)/2$ independent entries $(C_n)_{ij} = -c_{ij}$. In this gauge, the integral simplifies to
\be
    \psi_n(\Lambda) = \int \prod_{i < j} \dif c_{ij} \: \delta(C \cdot \Lambda) \, A_n(c_{ij}) \,.
\ee
The gauge-fixed version of the integral will be useful in defining the spinor embedding of the orthogonal Grassmannian in Section~\ref{sec: spinor emb}.

\subsection{Grassmannian Correlators}
The main advantage of the integral representation \eqref{eq:masterint} is that the Grassmannian correlator $A_n(C)$ carries the same physical information as $\psi_n(\Lambda)$, while also being a much simpler object. For this reason, its properties will be the main target of this paper. Invariance under $\mathrm{GL}(n)$ transformations demands that the function $A_n(C)$ depend on $C$ through its minors
\be\label{equ:minors}
    (I_1 \cdots I_n) = \epsilon^{a_1 \dots a_n} C_{a_1 I_1} \cdots C_{a_n I_n} \,,
\ee
with $\epsilon^{a_1 \dots a_n}$ the Levi-Civita symbol. These are subject to homogeneous quadratic relations known as the \emph{Plücker relations}. This statement follows from the observation that minors transform as $(I_1 \cdots I_n) \mapsto \det(R) (I_1 \cdots I_n)$ for $C \mapsto R \cdot C$. In order for the integral \eqref{eq:masterint} to be well-defined projectively under a $\mathrm{GL}(n)$ transformation, it must hold that
\be
    \label{eq:A4constraint2}
    A_n((I_1 \cdots I_n)) \mapsto \det(R)^{-(n-3)} A_n((I_1 \cdots I_n)) \,.
\ee
Each external field with spin $\ell_i$ is entirely characterized by its momentum and its \textit{helicity} $h_i = \pm \ell_i$. Given the labeling of the columns of $C$ by $\bar 1, \dotsm, \bar n, 1, \dotsm,n,$ the information about the external helicities is encoded as follows. Under little group transformations, which are realized on $C$ via
\be
    \label{eq:LGactionC}
    C \mapsto C \cdot \rho^{-1}, \quad \rho = \mathrm{diag} \left(\rho_1, \dots, \rho_n, \frac{1}{\rho_1}, \dots, \frac{1}{\rho_n} \right),
\ee
with $\rho_i \in \mathbb{R}^*$, the correlator transforms as
\be
    \label{eq:A4constraint1}
    A_n((I_1 \cdots I_n)) \mapsto \left( \prod_{i=1}^n \rho_i^{-2h_i} \right) A_n((I_1 \cdots I_n)) \,.
\ee
For $n = 3$, the above constraints imply that a given helicity assignment always admits only two solutions, one for each branch. For external spin-$1$ fields, one solution corresponds to a Yang--Mills interaction in the bulk, while the second one describes a $F^3$ interaction~\cite{Baumann:2024ttn}. For $n=4$, on the other hand, any little group-invariant function can be expressed in terms of the following ``Mandelstam variables''
\be
    S = (\bar 1 \bar 2 1 2), \quad T = (\bar 1 \bar 4 1 4), \quad U = (\bar 1 \bar 3 1 3) \,.
\ee
Crucially, the sum of these Mandelstams is not zero, but the locus $S+T+U = 0$ is closely connected to the corresponding scattering amplitudes. As a concrete example, consider the correlator of four gluons with two positive and two negative helicities. The function $A_4^{--++}$ that satisfies \eqref{eq:A4constraint1} and \eqref{eq:A4constraint2} then takes the form\footnote{We are implicitly assuming that the correlators describing an even number of negative helicity gluons localize in the right branch, as otherwise it would be impossible to factor $(12\bar 3\bar 4)^2$, which vanishes in the left branch, out.}
\be
    A_4^{--++} = (12\bar 3 \bar 4)^2 \, \frac{\mathcal{P}(S,T,U)}{\mathcal{Q}(S,T,U)} \,,
\ee
where $\mathcal{Q}$ is a polynomial of degree $N+3$ and $\mathcal{P}$ is a polynomial of degree $N$, with $N$ unfixed at this stage. A change in the external helicities amounts to swapping barred and unbarred columns in $(12\bar3\bar4)^2$.

\vskip 4pt
In \cite{Arundine:2026fbr}, the \emph{tree-level} four-point expressions were constructed by demanding a consistent factorization into three-point correlators on the singular loci $S = 0$ and $T = 0$. On $S = 0$, for example, it must hold that
\be
    \label{eq:factorule}
    \mathop{\rm Res}_{S = 0} A_4 = \sum_h A_{3,L}^{(-h)} \hs A_{3,R}^{(+h)} \,,
\ee
where $A_{3,L}^{(-h)}$ and $A_{3,R}^{(+h)}$ are the relevant three-point correlators, and $h$ is the helicity of the exchanged particle. The formula must be understood in terms of the following identities for the minors of $\ogr(4,8)$:
\be\label{equ:minors-factorized}
\begin{aligned}
(L_1L_2R_1R_2)&\ \xrightarrow{\ S=0\ }\ (L_1L_2 I_s)(R_1R_2\bar I_s)+(L_1L_2\bar I_s)(R_1R_2 I_s) \,, \\
(L_1L_2L_3R_1)&\ \xrightarrow{\ S=0\ }\	-(L_1L_2L_3)(R_1 I_s\bar I_s)\,,\\
(L_1R_1R_2R_3)&\ \xrightarrow{\ S=0\ }\	(L_1 I_s\bar I_s)(R_1R_2R_3)\,,
\end{aligned}
\ee	
where $L_i=\bar1,\bar2, 1,2, \: R_i=\bar3,\bar4, 3,4$, and $\bar I_s,I_s$ are the columns associated with the exchanged particle. In the chart \eqref{eq:rightchart} for both $\ogr(3,6)$ and the resulting $\ogr(4,8)$, these identities simplify to the gluing rule
\be
    \label{eq:simplegluing}
    c_{ij} \xrightarrow{\ S=0\ } c_{is}^L \hs c_{sj}^R \quad \mathrm{for} \: i =1,2 \quad \mathrm{and} \quad j=3,4 \,.
\ee
Note that these parameters trivially satisfy $S = c_{13}c_{24}-c_{14}c_{23} = 0$ and are invariant under the little group transformations of the exchanged particle, $c_{is}^L \mapsto t c_{is}^L, \hs c_{sj}^R \mapsto t^{-1} c_{sj}^R$, which is then quotiented out.

\vskip 4pt
The Yang--Mills color-ordered correlator, stripped of the helicity factor, is~\cite{Arundine:2026fbr}
\be
    \label{eq:fullcolor}
    A_4 = \frac{3}{(S+T+U) S \, T} + \frac{1}{(S+T-U) S \, T} - \frac{1}{(-S+T+U) S \, T} - \frac{1}{(S-T+U) S \, T} \,.
\ee
One can also define the following ``reduced'' color-ordered correlator, again stripped of the helicity factor, which is more amenable to momentum-space computations:
\be
    \label{eq:YMredref}
    \hat A_4 = \frac{1}{2} \Big( A_4(U) + A_4(-U) \Big) = \frac{4 \hs (S+T)}{(S+T+U)(S+T-U) S \hs T} \,.
\ee
Similar reduced correlators can be defined for the other Mandelstams as well. In this paper, these answers will be shown to have a clear geometric interpretation inside the cosmological Grassmannian. In the next subsection, we describe how the Grassmannian representation of scattering amplitudes can be extracted from the associated cosmological functions.

\subsection{Flat-Space Limit}
For scattering processes involving $n$ gluons, $k$ of which have negative helicity, the following representation is known~\cite{Arkani-Hamed:2009hub, Arkani-Hamed:2009ljj}:
\beq
\label{eq:amplitudegrass}
    M_n(\mathcal{L}, \tilde{\mathcal{L}}) = \int \frac{\ud^{k \times n} {\cal C}}{\mathrm{GL}(k)} \, \delta({\cal C} \cdot \tilde{\mathcal{L}}) \, \delta({\cal C}^\perp \cdot \mathcal{L}) \, \mathcal{M}_n(\mathcal{C})\,,
\eeq
where the integral is over the Grassmannian $\gr(k,n)$, i.e.~the space of $k$-planes in $\mathbb{R}^n$. For $k=2$, for example, the \emph{tree-level} Yang--Mills integrand is
\be
    \mathcal{M}_n(\mathcal{C}) = \frac{(12)^4}{(12)(23) \cdots (n1)}\,,
\ee
where $(i \,i+1)$ denotes the determinant of the columns $i$ and $i+1$ of the $2 \times n$ matrix $\mathcal{C}$. The $n \times 2$ matrices $\mathcal{L}$ and $\tilde{\mathcal{L}}$ contain the flat-space spinors $\lambda_i^\alpha$ and $\tilde \lambda_i^{\dot \alpha}$, with $\alpha,\dot\alpha = 1,2$, respectively, and four-momentum conservation implies $\mathcal{L}^T \cdot \tilde{\mathcal{L}} = 0$. The matrix ${\cal C}^\perp$ is the $(n-k)$-dimensional complement of ${\cal C}$, defined by ${\cal C}^\perp \cdot {\cal C}^T = 0$.

\vskip 4pt
We will now identify a limit in which the cosmological Grassmannian $\mathrm{OGr}(n,2n)$ reduces to the flat-space Grassmannian  $\mathrm{Gr}(k,n)$. To this end, it is useful to consider the following parametrization:
\be \label{equ:parametr-n-k}
C=\begin{pmatrix}- c_{ij}& 0_{k\times (n-k)}& 1_{k\times k}&-c_{iJ}\\-c_{Ij}& 1_{(n-k)\times (n-k)} & 0_{(n-k)\times k}&-c_{IJ}\end{pmatrix} ,
\ee 
where we have set the matrix with columns $1,2,\dots,k,\overline{k+1},\overline{k+2},\dots,\bar n$ to be the identity. This chart parameterizes the right or left branch of $\OGr(n,2n)$ if $k$ is even or odd, respectively.  We have split the
non-trivial components of the Grassmannian into two skew-symmetric matrices $c_{ij}$ and $c_{IJ}$, with $i,j=1,2,\dots,k$ and $I,J=k+1,\dots,n$, as well as a rectangular matrix~$c_{iJ}$, with $c_{Ji}=-c_{iJ}$.
Taking the limit $c_{ij},c_{IJ}\to0$, the structure of $C$ simplifies to
\be
    C \rightarrow C_{\rm flat} = \begin{pmatrix} 0_{k\times n} & \mathcal{C}_{k \times n}\\ \mathcal{C}^\perp_{(n-k) \times n} & 0_{(n-k)\times n}\end{pmatrix} ,
\ee
where we read off a gauge-fixed representative of $\mathcal{C}$ and $\mathcal{C}^\perp$ appearing in \eqref{eq:amplitudegrass}.
As a consequence of this limit, the Grassmannian Yang--Mills correlators are expected to satisfy the following schematic property, as proven in~\cite{Arundine:2026fbr}:
\be
    \mathop{\mathrm{Res}}_{c_{ij} = c_{IJ} = 0} A_n(c_{iJ},c_{ij},c_{IJ}) \propto \mathcal{M}_n(c_{iJ}) \,.
\ee
For $n= 4$, requiring this property and the connection with the non-negative Grassmannian $\gr(2,4)_{\geq 0}$ will be crucial in identifying the positive region $R$ inside $\OGr_R(4,8)$.

\subsection{Pfaffian Coordinates}\label{sec: spinor emb}
In the previous subsections, we have described an element of $\OGr_R(n,2n)$ as a gauge-fixed matrix $C=\left(1_{n \times n}, \, C_n\right)$, where $C_n$ is skew-symmetric. We have also discussed a parametrization in terms of its minors, which furnishes a chart-independent characterization. Either option, however, comes at a price. The gauge-fixed matrix only parametrizes a subset of the variety, while the minors are an extremely redundant language.

\vskip 4pt
For this reason, in this subsection we describe an embedding of $\OGr(n,2n)$ that covers the entirety of a branch, while also minimizing the number of necessary variables. From now on, $\mathrm{(O)Gr}(k,n)$ will always refer to the complexification of the real Grassmannian defined earlier.
Consider the rational map $\varphi: \OGr_R(n,2n)  \dashrightarrow \mathbb{P}^{2^{n-1}-1}$ given by
\be
    \label{eq:spinorembed}
    \left(1_{n \times n}, \, C_n\right) \mapsto \left(\Pf_I(-C_n): I \subset [n] \text{ of even size} \right),
\ee
where $[n]\equiv \{1,2,\cdots,n\}$, $\Pf_I(-C_n)$ is the Pfaffian of the principal submatrix of $-C_n$ given by selecting the rows and columns in $I$, and we set $\Pf_\emptyset(-C_n)\equiv 1$. The map $\varphi$ extends to all of $\OGr_R(n,2n)$, giving rise to its \emph{spinor embedding}. We denote, for $I \subset [n]$ even, the \textit{Pfaffian coordinates} in $\mathbb{P}^{2^{n-1}-1}$ by $p_I$.

\vskip 4pt
We now briefly recall the defining equation of Pfaffians and the relations between them. The determinant $\det(A)$ of a skew-symmetric matrix $A = (a_{ij})_{i,j=1}^{2m}$ of size $2m \times 2m$ is a homogeneous polynomial of degree $2m$ in the entries of $A$. Moreover, it is the square of a degree $m$ homogeneous polynomial $\Pf(A)$ known as the \emph{Pfaffian}:
    \begin{equation}\label{eq: Pfaffian}
        \Pf(A) \equiv\frac{1}{2^m \hs m!} \sum_{\sigma \in S_{2m}} {\rm sgn}(\sigma) \prod_{i=1}^{m} a_{\sigma(2i-1),\sigma(2i)} \,.
    \end{equation}
Analogously to the Plücker relations for determinants, these polynomials satisfy \emph{Pfaffian relations}. Let $I=\{i_1,\ldots,i_r\}$ and $J=\{j_1,\ldots,j_s\}$ be subsets of $[2m]$, each containing an odd number of elements. It then follows that the Pfaffians of $A$ satisfy the following relation:
\begin{equation}\label{eq: pfaffian rels}
     \sum_{k=1}^s (-1)^k p_{I\cup \{j_k\}}\cdot p_{J \setminus \{j_k\}} + \sum_{l=1}^r (-1)^l p_{I \setminus \{i_l\}}\cdot p_{J\cup \{i_l\}}=0 \,,
\end{equation}
where $p_{I\cup \{j_k\}}$ and $p_{J\cup \{i_l\}}$ carry a sign given by concatenating $j_k,i_l$ at the end and sorting the indices. In $\mathbb{P}^{2^{n-1}-1}$, the image $\varphi(\OGr_R(n,2n))$ is cut out precisely by \eqref{eq: pfaffian rels}, where $I$ and $J$ vary over all odd subsets of $[n]$. We describe the origin of this embedding in more detail in Appendix~\ref{app: Spinors}.

\vskip 4pt
For $n=3$, there are no Pfaffian relations and the spinor embedding exhibits the isomorphism $\OGr_R(3,6) \cong \PP^3$ via
\be\label{eq:pfaffian_emb_3}
    \begin{pmatrix}
        1 & 0 & 0 & 0 & -c_{12} & -c_{13}\\
        0 & 1 & 0 & c_{12} & 0 & -c_{23}\\
        0 & 0 & 1 & c_{13} & c_{23} & 0
    \end{pmatrix} \mapsto (1: c_{12}: c_{13}: c_{23}) \,.
\ee
For $n=4$, the corresponding rational map is
\be
    \label{eq:fourpointmap}
    \begin{pmatrix}
        1 & 0 & 0 & 0 &0 & -c_{12} & -c_{13} & -c_{14}\\
        0 & 1 & 0 & 0 & c_{12} & 0 & -c_{23} & -c_{24}\\
        0 & 0 & 1 & 0 &  c_{13} & c_{23} & 0 & -c_{34}\\
        0 & 0 & 0 & 1 & c_{14} & c_{24} & c_{34} & 0
    \end{pmatrix}\mapsto (1:c_{12}:c_{13}: c_{14}:c_{23}:c_{24}:c_{34}: c_{1234}) \,,
\ee
with $c_{1234} =c_{12}c_{34}-c_{13}c_{24}+c_{14}c_{23}$. This map extends to an embedding of $\OGr_R(4,8)$ onto a quadric hypersurface in $\PP^7$, defined as the locus
\be
p \hs p_{1234} - p_{12} \hs p_{34} + p_{13} \hs p_{24} - p_{14} \hs p_{23} =0\,,
\ee
where the set $\{ p,p_{12},p_{13}, p_{14},p_{23},p_{24},p_{34},p_{1234} \}$ denotes the homogeneous coordinates of $\PP^7$.

\newpage
\section{Positive Geometries}
\label{sec:positive_geometries}
In this section, we introduce the notion of positive geometries and define their properties. We start by recalling the more intuitive definition of positive geometries from \cite{Arkani-Hamed:2017tmz}. We then move on to the novel framework of positive geometries proposed by Brown and Dupont in \cite{brown2025positivegeometriescanonicalforms}, which is based on mixed Hodge theory of relative homology groups.
This is the point of view that will be most helpful in our study of the Grassmannian correlators in Section~\ref{sec:YM_from_geometry}. This section features several explicit examples, which help build intuition behind the more formal elements of the discussion.

\subsection{Motivation and Definition}\label{sec:motivation_PG}
Positive geometries are essentially semi-algebraic sets, i.e.~sets defined by polynomial equations and/or inequalities, which come equipped with a unique differential form that exhibits logarithmic poles along its boundaries and whose residues satisfy a recursive property.

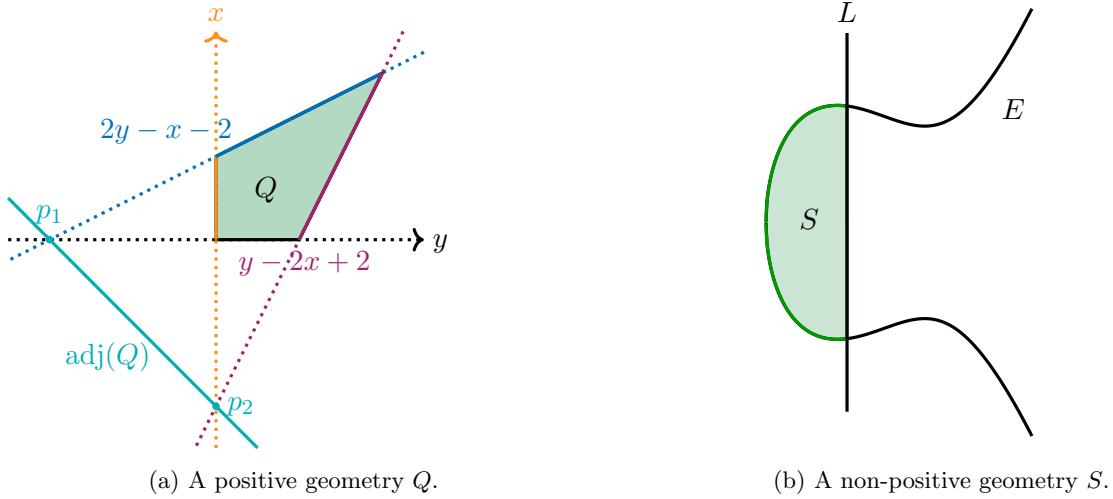
\begin{figure}
    \centering
    \begin{subfigure}{0.47\textwidth}
        \begin{tikzpicture}[scale=1.1]
    
            \draw[very thick, fill=ForestGreen!30] (0,1) -- (2,2) -- (1,0) -- (0,0) -- cycle;
            \node (Q) at (0.6,0.6) {$Q$};
          
            \draw[very thick,->, dotted] (-2.5,0) -- (2.5,0);
            \node at (2.7,-0.05){$y$};
            \draw[very thick,-, BurntOrange] (0,0) -- (0,1);
            \draw[very thick,->, BurntOrange, dotted] (0,-2.5) -- (0,2.5); 
            \node at (0.0, 2.7){\textcolor{BurntOrange}{$x$}};
          
            \draw[very thick,TealBlue] (-2.5,0.5) -- (0.5,-2.5);
            \node (adj) at (-1.3,-1.4) {\textcolor{TealBlue}{$\operatorname{adj}(Q)$}};

            \draw[very thick, RoyalBlue] (0,1) -- (2,2);
            \node[] () at (-0.6,1.3) {\textcolor{RoyalBlue}{$2y-x-2$}};
            \draw[very thick, dotted, RoyalBlue] (2.5,2.25) -- (-2.5,-0.25);
      
            \draw[very thick, RedViolet] (1,0) -- (2,2) node[pos = 0, below, xshift=2pt] {$y-{2}x+2$};
            \draw[very thick, dotted, RedViolet] (2.25,2.5) -- (-0.25,-2.5);
      
      
            \node[circle, fill=TealBlue, inner sep = 1] () at (-2,0) {};
            \node[inner sep = 1] () at (-2,0.3) {\textcolor{TealBlue}{$p_1$}};
      
            \node[circle, fill=TealBlue, inner sep = 1] () at (0,-2) {};
            \node[inner sep = 1] () at (0.3,-2) {\textcolor{TealBlue}{$p_2$}};
        \end{tikzpicture}
        \caption{A positive geometry $Q$.}
        \label{fig:Q_pos_geom}
    \end{subfigure}
    \hfill
    \begin{subfigure}{0.47\textwidth}
    \begin{tikzpicture}[scale=1, >=stealth]

        \pgfmathsetmacro{\xL}{-0.45}

        \pgfmathsetmacro{\xLeft}{-1.52138}

        \pgfmathsetmacro{\yL}{sqrt(\xL*\xL*\xL - \xL + 2)}

        \begin{scope}
            \clip (-3,-3) rectangle (\xL,3);

            \fill[ForestGreen!20]
                plot[domain=\xLeft:2, samples=300, smooth]
                    (\x,{sqrt(max(0,\x*\x*\x - \x + 2))})
                --
                plot[domain=2:\xLeft, samples=300, smooth]
                    (\x,{-sqrt(max(0,\x*\x*\x - \x + 2))})
                -- cycle;
        \end{scope}

        \draw[very thick, domain=\xLeft:2, samples=300, smooth]
            plot (\x,{sqrt(max(0,\x*\x*\x - \x + 2))});

        \draw[very thick, domain=\xLeft:2, samples=300, smooth]
            plot (\x,{-sqrt(max(0,\x*\x*\x - \x + 2))});

        \draw[
            very thick,
            green!60!black,
            line cap=round,
            line join=round
        ]
            plot[domain=\xLeft:\xL, samples=300, smooth]
                (\x,{sqrt(max(0,\x*\x*\x - \x + 2))})
            --
            plot[domain=\xL:\xLeft, samples=300, smooth]
                (\x,{-sqrt(max(0,\x*\x*\x - \x + 2))});

        \draw[very thick] (\xL,-2.5) -- (\xL,2.5)
            node[above] {$L$};

        \node at (1.75,1.5) {$E$};
        \node at (-0.95,0.05) {$S$};

    \end{tikzpicture}
    \caption{A non-positive geometry $S$.}
    \label{fig:E_non_pos_geom}
    \end{subfigure}
    \caption{Example and non-example of positive geometries. The lines in the left picture are labeled by their vanishing locus.}
\end{figure}

\vskip 4pt
It is best to start with some motivating examples.
Let us first consider the green-shaded (real) quadrilateral $Q\subset \PP^2$ in Figure~\ref{fig:Q_pos_geom}.
Denote the coordinates of the affine patch $U_z=\CC^2\subset \PP^2$ by $x$ and $y$.
The facets of the quadrilateral $Q$ are supported by the vanishing of the four lines
\be
    \begin{array}{l}
         \textcolor{Black}{L_1 = y} \,, \\[4pt]
         \textcolor{BurntOrange}{L_2 = x} \,, \\[4pt]
         \textcolor{NavyBlue}{L_3 = 2y-x-2} \,, \\[4pt]
         \textcolor{RedViolet}{L_4 =y-2x+2} \,.
    \end{array}
\ee
The algebraic boundary $\partial_aQ$ of $Q$ is therefore given by
\be
     x \hs y(2y-x-2)(y-2x+2)=0 \,.
\ee
We denote by $\operatorname{adj}(Q)$ the line $2(y+x+2) = 0$ interpolating the points $p_1$ and $p_2$, which are the intersections of the $4$ lines supporting the facets of $Q$ that do not lie in $Q$.
We can then write the unique canonical form of $Q$ (up to sign) as
\be\label{eq:CF_Q}
    \omega_Q = \frac{\operatorname{adj}(Q)}{\partial_aQ}\hs \dif x\wedge \dif y 
    =\frac{2(y+x+2)}{x \hs y(2y-x-2)(y-2x+2)} \hs \dif x \wedge \dif y \,.
\ee
Taking the residue along $x=0$ yields
\be
   \mathop{\Res}_{x=0} \omega_Q = \frac{1}{y (y-1)} \hs \dif y \,,
\ee
which is the canonical form of the orange line segment of $Q$.
In turn, the iterated residue of $\omega_Q$ at the origin equals $-1$, which is thought of as the canonical form of the zero-dimensional positive geometry, i.e.~the origin $(0,0)\in U_z$.
Note that the line $\operatorname{adj}(Q)$ precisely cancels the spurious poles of $\omega_Q$ at $p_1$ and $p_2$.

\vskip 4pt
Next, let us also briefly discuss a counterexample to the positive-geometry property, which was already examined in \cite[Section 5]{Arkani-Hamed:2017tmz}.
Consider the (real) region $S\subset \PP^2$ bounded by a smooth elliptic curve $E$, defined by $y^2z-x^3+xz^2-2z^3=0$, and a line $L$.
The region $S$ in the affine chart $U_z=\{z\neq 0\}\subset \PP^2$ is shown in Figure~\ref{fig:E_non_pos_geom}. Following the same logic as above, a contender for the canonical form $\omega_S$ would have poles along $E$ and $L$.
However, the elliptic curve $E$ supports a non-zero global holomorphic
$1$-form $\omega_E$. On the affine chart $U_z$, it can be written as
\be
    \omega_E
    = \frac{\dif x}{2y}
    = \frac{\dif y}{3x^2-1}\,.
\ee
Since $2y$ and $3x^2-1$ do not vanish simultaneously on $E$, at every point of $E$ there exists a regular representation of $\omega_E$, hence it is globally holomorphic. The form also
extends regularly to the unique point $(0:1:0) \in E$ outside $U_z$, as can be shown in terms of the coordinates $u = x/y$ and $v = z/y$ in the chart $y \neq 0$.

\vskip 4pt
Suppose we had a candidate for a canonical form $\omega_S$ on $S$. Its residue along $E = 0$ would then need to be the canonical form on the curve segment shown in \textcolor{ForestGreen}{dark green} in Figure~\ref{fig:E_non_pos_geom}.
However, note that the one-dimensional family of $1$-forms
\be
  \bigg( \mathop{\rm Res}_{E=0}\omega_S \bigg) + \lambda \omega_E \,,
\ee
with $\lambda \in \CC$, all have the correct residues along the boundaries of the line segment, since $\omega_E$ does not have any non-trivial residues.
As a consequence, the curve segment cannot be a positive geometry due to the lack of uniqueness of the forms at each recursive step, and by extension neither can $S$.

\begin{figure}
    \centering
    \begin{tikzpicture}[
        node distance=1.2cm and 1.5cm,
        every node/.style={inner sep=1pt},
        edge/.style={very thick}
    ]
    
    \node (Q) at (0,3) {$\PP^2$};
    
    \node (L1) at (-3,1.5) {$L_1$};
    \node (L2) at (-1,1.5) {$L_2$};
    \node (L3) at (1,1.5) {$L_3$};
    \node (L4) at (3,1.5) {$L_4$};
    
    \node[TealBlue] (r1) at (-3.5,0) {$p_1$};
    \node[TealBlue] (r2) at (3.5,0) {$p_2$};
    \node (v12) at (-2.5,0) {$v_1$};
    \node (v23) at (-0.8,0) {$v_2$};
    \node (v34) at (0.8,0) {$v_3$};
    \node (v41) at (2.5,0) {$v_4$};
    
    \node (empty) at (0,-1.5) {$\varnothing$};
    
    \draw[edge, Black!100] (Q) -- (L1);
    \draw[edge, Black!100] (Q) -- (L2);
    \draw[edge, Black!100] (Q) -- (L3);
    \draw[edge, Black!100] (Q) -- (L4);
    
    \draw[edge] (L1) -- (v12);
    \draw[edge] (L2) -- (v12);
    
    \draw[edge] (L2) -- (v23);
    \draw[edge] (L3) -- (v23);
    
    \draw[edge] (L3) -- (v34);
    \draw[edge] (L4) -- (v34);
    
    \draw[edge] (L4) -- (v41);
    \draw[edge] (L1) -- (v41);

    \draw[edge, TealBlue] (L1) -- (r1);
    \draw[edge, TealBlue] (L3) -- (r1);
    \draw[edge, TealBlue] (L2) -- (r2);
    \draw[edge, TealBlue] (L4) -- (r2);
    
    \draw[edge] (v12) -- (empty);
    \draw[edge] (v23) -- (empty);
    \draw[edge] (v34) -- (empty);
    \draw[edge] (v41) -- (empty);
    \draw[edge,TealBlue] (r1) -- (empty);
    \draw[edge,TealBlue] (r2) -- (empty);

    \end{tikzpicture}
    \caption{Boundary stratification $\mathcal{L}(\partial_aQ)$ of the quadrilateral $Q$ in Figure~\ref{fig:Q_pos_geom}. 
    The components are given by the lines $L_1,\ldots,L_4$ bounding $Q$, and the vertices $v_1,\ldots, v_4$ bounding the associated line segments.
    The components in \textcolor{TealBlue}{teal} are the residual arrangement of $Q$, which we define formally in Section~\ref{sec:residual_arrangement}.}
    \label{fig:bd_strat_Q}
\end{figure}
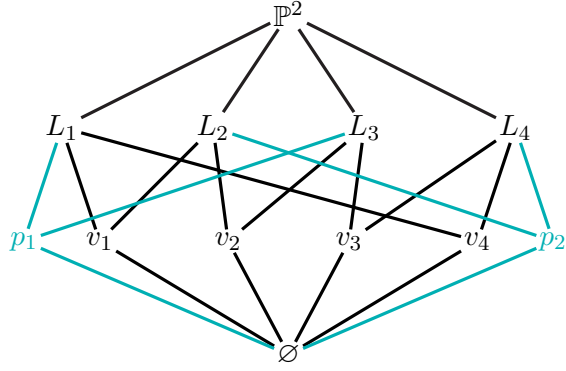

\vskip 4pt
We now move on to a more mathematical description of positive geometries.
We start by fixing some notation.
Let $X\subset \PP^N$ be a projective complex $n$-dimensional variety.
Denote by $X(\RR)$ the set of real points in $X$.
Let $R\subset X(\mathbb{R})$ be an orientable top-dimensional semi-algebraic set in the real points of $X$.
Moreover, let $R$ be \emph{regular}, that is, the Euclidean closure of its interior is again $R$.
Let $D = \partial_a R$ be the \emph{algebraic boundary} of $R$, defined as the Zariski closure of the Euclidean boundary $\partial R$, that is, the smallest subvariety of $X$ containing $\partial R$.
Suppose $D$ has irreducible components $D_1,\ldots, D_m$.
We then collect all the irreducible components of the intersections $D_I=\bigcap_{i\in I}D_i$, for all $I\subset \{1,\ldots,m\}$, into a partially ordered set (poset) $\mathcal{L}(D)$, called the \emph{boundary stratification}. The ordering in this set is defined by inclusion, i.e.~$A \prec B$ if $A \subset B$.
The boundary stratification associated with the quadrilateral $Q$ in Figure~\ref{fig:Q_pos_geom}, where its vertices are denoted by $v_1,\ldots, v_4$, is given in Figure~\ref{fig:bd_strat_Q}.

\vskip 4pt
We move on to the definition of canonical forms. Suppose that, in local coordinates $x_1,\ldots, x_n$ on $X$, we can write $D = \{x_1\cdots x_r=0\}$, with $r\leq n$. 
A differential $k$-form $\omega_k$ on $X \!\setminus\! D$ is then an \emph{algebraic $k$-form with logarithmic poles along $D$} if it can be written as a linear combination of 
\be
  \omega_k^{(p)}  = \alpha_q \wedge \frac{\dif x_{i_1}}{x_{i_1}}\wedge \ldots \wedge \frac{\dif x_{i_p}}{x_{i_p}} \,,
\ee
where $\alpha_q$ is a (locally) holomorphic $q$-form on $X$, the indices range in $1\leq i_1<\ldots < i_p\leq r$, and $p+q=k$.
The $\CC$-vector space of algebraic $k$-forms with logarithmic poles along $D$ is denoted by $\Omega^k_{\log}(X\setminus D)$.
Let $D_i$, with $i=1,\ldots,m$, be the irreducible components of $D$.
If $X\!\setminus\! D$ is smooth, then there is a well-defined \emph{Poincar\'e residue} map along $D_i$
\be
   \mathrm{Res}_{D_i}: \Omega^k_{\log}(X\!\setminus\! D) \rightarrow \Omega^{k-1}_{\log}(D_i \!\setminus\! D'_i) \,,
\ee
where $D'_i \equiv D_i\cap \big(\bigcup_{j\neq i}D_j \big)$, which sends
\be
    \label{eq:residuemap}
    \omega_k = \frac{\dif D_i}{D_i}\wedge \eta + \zeta \mapsto \eta\vert_{D_i} \,,
\ee
where $\eta$ and $\zeta$ are a $(k{-}1)$-form and a $k$-form, respectively, which do not have poles on $D_i$.

\vskip 4pt
Equipped with the above notions, we can finally give a proper definition of a positive geometry following that of \cite{Arkani-Hamed:2017tmz}, which we record here for ease of reference.
\begin{definition}\label{def:rec_PG}
     A \emph{(recursive) positive geometry} is a triple $(X,R,\omega_R)$, consisting of the $n$-dimensional ambient variety $X\subset\PP^N$, the top-dimensional orientable semi-algebraic set $R$ with algebraic boundary $D$, and a differential top-form $\omega_R$, called \emph{canonical form}, such that 
\begin{enumerate}
        \item For $n>0$, the canonical form is unique and $\omega_R\in \Omega^n_{\log}(X\!\setminus\! D)$.
        \item For each irreducible component $D_i$, the triple $(D_i,R_i, \omega_{R_i})$ is a positive geometry, where $\omega_{R_i}$ is the Poincaré residue $\operatorname{Res}_{D_i} \omega_R$ and $R_i$ is the closure of the Euclidean interior \linebreak $\operatorname{int}(R \cap D_i(\RR))$ in $D_i(\RR)$, with orientation induced by that of $\operatorname{int}(R)$. 
        \item If $n=0$, then the ambient variety $X=R$ is a point and $\omega_R = \pm 1$. 
        The sign determines the orientation on this zero-dimensional manifold.
    \end{enumerate}
\end{definition}

\noindent
For example, following this definition, the triple $(\PP^2,Q,\omega_Q)$ for the quadrilateral $Q$ in Figure~\ref{fig:Q_pos_geom} is a positive geometry.

\subsection{Residual Arrangements and Adjoints}\label{sec:residual_arrangement}
Some of the main tools to prove the existence of positive geometries and compute canonical forms are the residual arrangement and the adjoint hypersurface.
We will now briefly introduce them.

\vskip 4pt
Let $X$ be a smooth projective variety, $R\subset X(\R)$ be a semi-algebraic set in the real points of $X$, and let $D\subset X$ be the algebraic boundary of $R$.
An irreducible component $Z\in\mathcal{L}(D)$ is called \emph{residual} if the dimension of the real semi-algebraic set $Z\cap R$ is strictly smaller than the dimension of $Z$ as a complex variety.
Put in symbols, $Z$ is residual if
\be\label{eq:def_residual}
    \dim_\mathbb{R}(Z\cap R)< \dim_\mathbb{C}(Z) \,.
\ee
For example, the points $p_1$ and $p_2$ in Figure~\ref{fig:Q_pos_geom} are residual, as they are $0$-dimensional varieties, but $p_i\cap Q=\varnothing$, for $i=1,2$, has negative dimension as a semi-algebraic set by convention.
The \emph{residual arrangement} $\mathcal{R}(R)\subset \mathcal{L}(D)$ is then the collection of residual components.
The residual arrangement of the quadrilateral $Q$ is shown in \textcolor{TealBlue}{teal} in Figure~\ref{fig:bd_strat_Q}.
Note that this is a refinement of the notion of residual arrangement that appears elsewhere in the literature; see e.g.~\cite{Ranestad_2024, koefler2025takingamplituhedronlimit}.

\vskip 4pt
The \emph{adjoint} $\operatorname{adj}(R)$ is the lowest degree polynomial defining the hypersurface whose vanishing contains the residual arrangement.
For the purposes of this work, we will be interested in adjoints on a smooth hypersurface inside a projective space. Let $X\subset \PP^{n}$ be a smooth $(n{-}1)$-dimensional hypersurface of degree $d$, i.e.~its defining equation is of degree $d$, and $D\subset X$ be the algebraic boundary of $R$.
We then have
\begin{align}
    \label{eq:deg_adj}
    \deg \operatorname{adj}(R) = \deg D + d - n-1 \,.
\end{align}
This ensures that $\omega_{n-1} \in\Omega^{n-1}_{\log}(X\!\setminus\! D)$ is projectively well-defined, as we now argue. Let $(x_0: x_1: \cdots : x_n)$ be the homogeneous coordinates on $\mathbb{P}^n$, and define the homogeneous $n$-form on $\PP^n$
\be
    \label{eq:projform}
    \omega_{\mathbb{P}^n} = \sum_{i = 0}^n (-1)^i \hs x_i \hs \dif x_0 \wedge \cdots \wedge \widehat{\dif x_i} \wedge \cdots \wedge \dif x_n,
\ee
where the hat means omission, which has weight $n+1$ under a projective rescaling of the coordinates.
Equivalently, given a rescaling $x \mapsto tx$, with $x \in \mathbb{P}^n$ and $t \in \mathbb{C}^*$, we have $\omega_{\mathbb{P}^n} \mapsto t^{n+1} \hs \omega_{\mathbb{P}^n}$.
A regular $(n{-}1)$-form $\omega_{n-1}$ on $X\!\setminus\! D$ is then given by the residue of $\omega_{\mathbb{P}^n}/g$ along $g = 0$, with $g$ the defining equation of $X$. Since $X$ is of degree $d$, the form $\omega_{\mathbb{P}^n}/g$ has weight $n+1-d$.
However, we are interested in forms which also have poles along $D$, which is of degree $\deg D$.
Hence, if $\omega_{\mathbb{P}^n}/g$ has weight $n+1-d$ and the extra denominator in $\omega_{n-1}$ has weight $\deg D$, then its numerator must scale as in \eqref{eq:deg_adj}. If $X = \mathbb{P}^n \subset \mathbb{P}^{n+1}$, the formula \eqref{eq:deg_adj} holds with $d = 1$.
Note that this degree count can be extended to any smooth subvariety $X$ of higher codimension, but this involves more algebro-geometric language and we will not need this extension here.

\vskip 4pt
For the quadrilateral $Q\subset \P^2$ in Figure~\ref{fig:Q_pos_geom}, the adjoint $\operatorname{adj}(Q)$ is of degree
\be
    \deg\operatorname{adj}(Q) = \deg \partial_aQ - 3 = 1 \,.
\ee
Indeed, $\operatorname{adj}(Q)$ is the \textcolor{TealBlue}{teal} line shown in the figure, whose equation is given by $\operatorname{adj}(Q) = y+x+2$.

\subsection{Mixed Hodge Theory}\label{subsec:mixedhodge}
Recently, Brown and Dupont \cite{brown2025positivegeometriescanonicalforms} suggested a new framework for positive geometries and canonical forms, based on the mixed Hodge structure of the relative homology groups of a pair of varieties $(X,D)$, with $D \subset X$. Here, $D$ plays the role of the algebraic boundary of a semi-algebraic set $R$ and $X$ plays the role of the ambient space.
They introduce two Hodge-theoretic invariants for a pair: the \emph{genus} $g(X,D)$ and the \emph{combinatorial rank} $\operatorname{cr}(X,D)$.
The genus essentially captures the number of independent global holomorphic forms on $X \!\setminus\! D$, that is, those forms which do not have any poles on $X\!\setminus\! D$.
A vanishing genus then guarantees that a region in the real locus of $X\!\setminus\! D$ can be assigned a unique top-differential form with logarithmic poles along $D$.
The combinatorial rank, on the other hand, is the dimension of $\Omega^{\dim X}_{\log}(X\!\setminus\! D)$ if $g(X,D)=0$. Hence, a non-vanishing combinatorial rank addresses the existence of a canonical form.

\vskip 4pt
We introduce the essential notions; see \cite{hatcher2002algebraic} for a detailed introduction.
Let $X$ be a complex projective variety of dimension $n$.
Denote by $H_k(X)$ the \emph{$k$-th singular homology group}, with coefficients in $\Q$, and denote the dual vector space $H^k(X)=\Hom(H_k(X),\Q)$ as the \emph{$k$-th singular cohomology group}.
If $X$ is smooth, then we can essentially identify the complexification $H^k(X)_{\CC}$ as the space of closed algebraic $k$-forms modulo the exact $k$-forms on $X$.
Let $D \subset X$ be a subvariety of $X$; then, we denote by $H_k(X,D)$ the \emph{k-th relative singular homology group}. Its elements, called relative $k$-cycles, are equivalence classes of singular $k$-simplices, i.e.~continuous maps $\sigma^k:\Delta^k\to X$ from the $k$-simplex $\Delta^k\subset \RR^k$, with boundary on $D$.
Similarly, we denote by $H^k(X,D)$ the dual vector space $\Hom(H_k(X,D),\Q)$, called the \emph{$k$-th relative cohomology group}.

\vskip 4pt
We now move on to Hodge structures. To get an intuition, we start with the simpler case of the Hodge structure on (the cohomology of) a smooth projective variety $X$.
In this setting, there is a natural decomposition of the cohomology groups on $X$, first identified by Hodge in~\cite{Hodge}:
\begin{align}\label{eq:pure_HS_decomp}
    H^k(X)_{\CC}=\bigoplus_{p+q=k}H^{p,q}(X) \,,
\end{align}
where, loosely speaking, $H^{p,q}(X)$ is the space of $(p,q)$-differential forms on $X$, i.e.~the space of forms which can be written locally as a holomorphic $p$-form times an anti-holomorphic $q$-form.
This is the \emph{Hodge decomposition}, and it gives $X$, or rather its cohomology groups, a \emph{pure Hodge structure of weight~$k$}.
The dimensions of the complex vector spaces $H^{p,q}(X)$ are the (pure) Hodge numbers
\be
\label{eq:Hodgenum}
h^{p,q}(X) = \dim H^{p,q}(X) = h^{q,p}(X) \,,
\ee
where the symmetry follows from complex conjugation.
We can organize them into a Hodge diamond.
For example, let $E$ be the smooth elliptic curve in Figure~\ref{fig:E_non_pos_geom}.
The complete Hodge diamond is then given in Figure~\ref{fig:hodge_diamond_E}.
Note that $h^{1,0}(E) =1$ detects precisely the existence of the global holomorphic $1$-form $\omega_E$ on $E$, which we encountered earlier in Section~\ref{sec:motivation_PG}.

\begin{figure}
\centering
\begin{tikzpicture}[scale=1.2]
\node at (0,2) {$h^{1,1}(E)=1$};

\node at (-1,1) {$h^{1,0}(E)=1$};
\node at (1,1) {$h^{0,1}(E)=1$};

\node at (0,0) {$h^{0,0}(E)=1$};

\end{tikzpicture}
\caption{Hodge diamond of an elliptic curve $E$.}
\label{fig:hodge_diamond_E}
\end{figure}

\vskip 4pt
In order to define the crucial invariants of this framework, the genus and the combinatorial rank, we need to introduce mixed Hodge structures. A mixed Hodge structure on the relative cohomology groups of $(X,D)$ essentially combines multiple pure Hodge structures together.
It can be realized as a bi-filtration on the relative cohomology groups $H^k(X,D)$; see \cite{Deligne1971TheorieHodgeII}.
The \emph{mixed Hodge numbers} are then the complex dimensions of the subquotient vector spaces of $H^k(X,D)$ induced by this bi-filtration.
For the middle cohomology groups $H^n(X,D)$, we write the Hodge numbers as $h^{p,q}(X,D)$.
By convention, we denote the mixed Hodge numbers on the middle relative homology group by $h^{-p,-q}(X,D)$, with $p,q\geq 0$, and we have $h^{p,q}(X,D) = h^{-p,-q}(X,D)$ by definition.
Note, in particular, that mixed Hodge structures can have $h^{p,q}(X,D)\neq 0$ with $p+q\neq n$, as opposed to a pure Hodge structure.

\vskip 4pt

The \emph{genus} of a pair $(X,D)$ is then defined as
\begin{align}\label{eq:genus_def}
g(X,D)\equiv \sum_{p>0}h^{-p,0}(X,D) \,.
\end{align}
Similarly, the \emph{combinatorial rank} of the pair $(X,D)$ is defined as
\begin{align}\label{eq:def_cr}
\operatorname{cr}(X,D)\equiv h^{0,0}(X,D) \,.
\end{align}
If $g(X,D)=0$, the combinatorial rank is precisely the dimension of $\Omega^n_{\log}(X\!\setminus\! D)$, the space of regular algebraic forms on $X\!\setminus\! D$ with logarithmic poles along $D$.
Note that if $X$ is smooth and $D=\varnothing$, then the mixed Hodge numbers recover the (pure) Hodge numbers in \eqref{eq:Hodgenum}; for example, $h^{p,q}(X,D)=h^{p,q}(X)$ with $p+q=n$.
Moreover, in this case, $g(X,\varnothing)=g(X)$ is the geometric genus of the smooth variety $X$, that is, the dimension of the space of global holomorphic top-dimensional forms on $X$.
For example, the existence of such a form on the elliptic curve $E$ from Figure~\ref{fig:E_non_pos_geom} precisely corresponds to the fact that $g(E,\varnothing) = g(E) =h^{1,0}(E,\varnothing)=1$.

\vskip 4pt
As an example of an application of this framework, consider the quadrilateral $Q\subset \PP^2$ in Figure~\ref{fig:Q_pos_geom} together with its algebraic boundary $\partial_a Q$, consisting of the union of the four lines supporting its facets.
Note that the line at infinity in Figure~\ref{fig:Q_pos_geom} is not part of the boundary. The region $Q$ defines a real relative $2$-cycle in $H_2(\PP^2,\partial_a Q)$.
The mixed Hodge structure on the relative homology of hyperplane arrangements is particularly simple.
In this case, all Hodge numbers with $p\neq q$ vanish, therefore $g(\PP^2,\partial_a Q)=0$; see e.g.~\cite[Lemma 3]{brieskorn1973groupesTresses}.
Moreover, by \cite[Proposition 6.2]{brown2025positivegeometriescanonicalforms}, we have $\operatorname{cr}(\PP^2,\partial_a Q)=3$, which equals the number of bounded regions in Figure~\ref{fig:Q_pos_geom}.

\vskip 4pt
If $g(X,D)=0$, one of the main results in \cite{brown2025positivegeometriescanonicalforms} is that there exists a well-defined \emph{canonical map}
\begin{align}
\label{eq:canonical_map}
    \Omega:H_n(X,D)\longrightarrow \Omega^n_{\log}(X\!\setminus\! D) \,,
\end{align}
which assigns any cycle $\sigma \in H_n(X,D)$ a canonical form $\Omega(\sigma)\in \Omega^n_{\log}(X\!\setminus\! D)$.
We will not give the details of this map here. 
Suffice it to say that, in the examples where both approaches are applicable, the canonical form $\Omega(\sigma)$ is precisely the canonical form that satisfies the previous definition of positive geometries.
For example, this is the case for complements of hyperplane arrangements.
We can now turn to the definition of a positive geometry using this framework.

\begin{definition}\label{def:MHS_PG}
Let $X$ be an orientable complex projective variety.
Let $D\subset X$ be a subvariety of codimension $1$, such that $X\!\setminus\! D$ is smooth.
Denote by $\iota:(X(\R),D(\R))\hookrightarrow (X,D)$ the natural inclusion. $(X,D)$ is a \emph{positive pair} if the following conditions hold:
\begin{enumerate}
    \item for all $Z\in\mathcal{L}(D)$, we have $g(Z,D_Z)=0$, where $D_Z$ is the union of all irreducible components in $\mathcal{L}(D)$ of codimension $1$ in $Z$.
    \item there exists a cycle $\sigma\in H = \iota_* H_n(X(\RR),D(\RR)) \subset H_n(X,D)$ which has a non-zero canonical form, i.e.~$\Omega(\sigma)\neq 0$;
\end{enumerate} 
Then, any real $n$-cycle $\sigma\in H$ such that $\Omega(\sigma) \neq 0$ is called a \emph{positive geometry}.
\end{definition}

Note that the first requirement above is the analogue of the recursive residue condition of the canonical form introduced previously, which guarantees the uniqueness of the form in each recursive step.
However, this definition allows for different values of the residues at the vertices, other than $\pm 1$.
Also note that this requirement is stronger than the one in \cite[Definition 5.4]{sturmfels2026positivegeometriescubicsurfaces}.

\vskip 4pt
We continue the example of the quadrilateral $Q$ from Figure~\ref{fig:Q_pos_geom}.
First, note that all subvarieties in the boundary stratification $\mathcal{L}(\partial_aQ)$ shown in Figure~\ref{fig:bd_strat_Q} are projective spaces.
Therefore, $g(Z)=0$ for all $Z\in\mathcal{L}(\partial_aQ)$.
It follows from \cite[Corollary 3.14]{brown2025positivegeometriescanonicalforms} that $g(Z,\partial_a Q_Z)=0$ for all $Z\in \mathcal{L}(\partial_aQ)$, where $\partial_a Q_Z$ consists of the codimension-$1$ components in $Z$.
This shows that $[Q]\in H_2(\PP^2,\partial_aQ)$ is a positive geometry as in Definition \ref{def:MHS_PG}, with canonical form
\be
    \Omega([Q]) = \omega_Q =\frac{2(y+x+2)}{x \hs y(2y-x-2)(y-2x+2)} \hs \dif x\wedge \dif y \,,
\ee
as in \eqref{eq:CF_Q}.

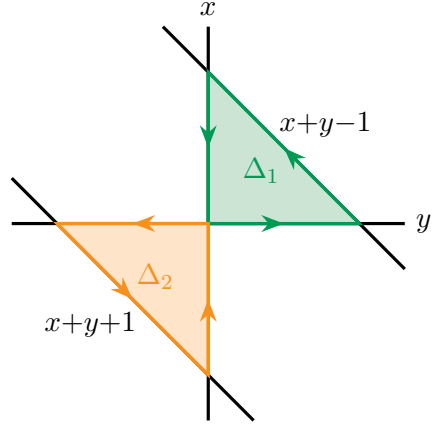
\begin{figure}
    \centering
     \begin{tikzpicture}[scale=2]

        \draw[-, very thick] (-1.3,0) -- (1.3,0) node[right] {$y$};
        \draw[-, very thick] (0,-1.3) -- (0,1.3) node[above] {$x$};
        \draw[-, very thick] (-0.3,1.3) -- (1.3,-0.3) node[pos=0.5, right, above, xshift =0.55cm] {$x{+}y{-}1$};
        \draw[-, very thick] (-1.3,0.3) -- (0.3,-1.3) node[pos=0.7, left, above, xshift =-1.2cm] {$x{+}y{+}1$};
        
        \fill[ForestGreen!20] (0,0) -- (1,0) -- (0,1) -- cycle;
        \draw[ForestGreen, very thick] (0,0) -- (1,0) -- (0,1) -- cycle;
        \draw[ForestGreen, very thick, midarrow] (0,0) -- (1,0);
        \draw[ForestGreen, very thick, midarrow] (1,0) -- (0,1);
        \draw[ForestGreen, very thick, midarrow] (0,1) -- (0,0);
        
        \fill[BurntOrange!20] (0,0) -- (-1,0) -- (0,-1) -- cycle;
        \draw[BurntOrange, very thick] (0,0) -- (-1,0) -- (0,-1) -- cycle;
        \draw[BurntOrange, very thick, midarrow] (0,0) -- (-1,0);
        \draw[BurntOrange, very thick, midarrow] (-1,0) -- (0,-1);
        \draw[BurntOrange, very thick, midarrow] (0,-1) -- (0,0);
        
        
        
        \node[ForestGreen] at (0.35,0.35) {$\Delta_1$};
        \node[BurntOrange] at (-0.35,-0.35) {$\Delta_2$};
        
        \end{tikzpicture}
    \caption{Two triangles in the plane connected in codimension $2$. The lines in the figure are labeled by their vanishing locus.}
    \label{fig:twoSimplices}
\end{figure}

\vskip 4pt
Finally, let us apply this framework to the two simplices meeting at the origin in Figure~\ref{fig:twoSimplices}, which serves as a toy model for the situation of interest in the orthogonal Grassmannian in Section \ref{sec:YM_from_geometry}.
Let $\mathcal{A}\subset \PP^2$ be the union of the four projective lines bounding $\Delta_1$ and $\Delta_2$.
We have $g(\PP^2,\mathcal{A})=0$ and, more generally, $g(Z,\mathcal{A}_Z)=0$ for all $Z \in \mathcal{L}(\mathcal{A})$ by the same reasoning as above.
Let $U_z\cong \CC^2$ be the chart $\{z \neq 0 \}$ on $\PP^2$, and choose an orientation on $\RR^2\subset U_z$, e.g.~the counter-clockwise orientation. The cycles $\Delta_1,\Delta_2\in H_2(\P^2,\mathcal{A})$ then inherit their orientation from that on $\RR^2$, as shown in Figure \ref{fig:twoSimplices}. 
We have, in particular:
\be
    \Omega([\Delta_1]) = \frac{1}{x \hs y(x+y-1)}\hs \dif x\wedge \dif y,\qquad \Omega([\Delta_2]) = \frac{-1}{x \hs y(x+y+1)}\hs \dif x\wedge \dif y \,.
\ee
Consider the union of these two simplices $\Delta = \Delta_1\cup\Delta_2$.
It is then natural in this framework to assign $[\Delta]$ the sum of the canonical forms of its two pieces, i.e. 
\be
    \Omega([\Delta]) = \Omega([\Delta_1])+\Omega([\Delta_2])=\frac{2}{x \hs y(x+y+1)(x+y-1)}\hs \dif x\wedge \dif y \,.
\ee
However, note that the residue at the origin equals $\pm2$ depending on the order of residues, and on the remaining vertices equals $\pm 1$.
This is in apparent contradiction to Definition \ref{def:rec_PG}.
We can resolve this by noting that the origin $(x:y:z)=(0:0:1)$ is a spurious pole of $\Delta$ and thus should not be considered a vertex, as recursively we obtain the full line segment $\Delta \cap \{x=0\}=[-1,1]$ as a boundary component.
Then, choosing orientations on the two simplices individually, which is possible since their interiors are disjoint, leads us to the following canonical form $\omega_{\Delta}$ of $\Delta$:
\be
    \omega_{\Delta} = \Omega([\Delta_1]-[\Delta_2])= \Omega([\Delta_1]) - \Omega([\Delta_2])=\frac{2(x+y)}{x \hs y(x+y+1)(x+y-1)}\hs \dif x\wedge \dif y \,,
\ee
whose residues at the vertices all equal $\pm1$.
Therefore, this gives rise to a positive geometry triple $(\PP^2,\Delta,\omega_{\Delta})$ as in Definition \ref{def:rec_PG}, as well as a positive geometry $[\Delta_1]-[\Delta_2]\in H_2(\PP^2,\mathcal{A})$ as in Definition \ref{def:MHS_PG}. 
Note that this choice of sign can be understood from the requirement that codimension-$k$ boundaries shared by two regions should cancel, taking into account their induced orientations. In our case, the regions in $\Delta$ meet along codimension-$2$ boundaries, hence the corresponding cycles must have relative sign $-(-1)^2=-1$.
Lastly, note that the numerator of the canonical form $\omega_\Delta$ defines the line through the origin and the intersection of the two parallel lines at infinity, hence it precisely cancels these spurious poles.

\newpage
\section{Yang--Mills from Geometry}\label{sec:YM_from_geometry}
The previous sections have set the stage to finally investigate positive geometries in the cosmological Grassmannian. In this section, we investigate three- and four-point correlators, and show explicitly how Yang--Mills theory emerges as the most natural theory that admits a positive-geometric picture in the Grassmannian at tree level. For $n=3$, the study of the Pfaffian-positive orthant of $\OGr_R(3,6)$ is straightforward. For $n=4$, on the other hand, we leverage the connection of the correlator with scattering amplitudes and lower-point data to determine it. In particular, we show how both the ``reduced'' and the ``full'' color-ordered Yang--Mills correlators can be obtained from our geometric construction.

\subsection{Three-Point Function}
We begin our study with the case $n=3$.
This case is straightforward, thanks to the isomorphism $\OGr_R(3,6)\cong \P^3$ shown in \eqref{eq:pfaffian_emb_3}. We want to identify a subregion $R$ with natural positivity properties and characterize it. The most natural candidate is the orthant $\PP^3_{\geq 0}$.

\vskip 4pt
Denote the coordinates of $\PP^3$ as $(p: p_{12}: p_{13}: p_{23})$. The non-negative orthant is then characterized by $p, \hs p_{ij} \geq 0$. A putative canonical form is
\be
    \label{eq:pfaffian3pt}
    \omega_{\geq 0} = \frac{\omega_{\mathbb{P}^3}}{p \hs p_{12} \hs p_{13} \hs p_{23}} \; \xmapsto{p=1} \; \frac{1}{p_{12} \hs p_{13} \hs p_{23}} \hs \dif p_{12}\wedge \dif p_{13}\wedge \dif p_{23} \,,
\ee
where the homogeneous form $\omega_{\mathbb{P}^n}$ was defined in \eqref{eq:projform}. As we argue below, $(\PP^3,\PP^3_{\geq 0},\omega_{\geq 0})$ is a positive geometry according to both Definition~\ref{def:rec_PG} and~\ref{def:MHS_PG}. In the chart \eqref{eq:pfaffian_emb_3}, where $p_{ij}\mapsto c_{ij}$, the canonical form $\omega_{\geq 0}$ reproduces exactly the Yang--Mills three-point function~\cite{Baumann:2024ttn, Arundine:2026fbr}:
\be
    A_3 = \frac{1}{(1 \bar 1 2)(\bar 2 3 \bar 3)} = \frac{1}{c_{12} \hs c_{13} \hs c_{23}} \,.
\ee
More precisely, it is the helicity-stripped correlator.

\vskip 4pt
We now show how to recover the canonical form of $\gr(2,3)_{\geq 0}$, which encodes the flat-space Yang--Mills scattering amplitude, from $\omega_{\geq 0}$. The flat-space locus is characterized by $(\bar 1 \bar 2 \bar 3) = (1 2 3) = 0$, as shown in~\cite{Arundine:2026fbr}.
In terms of the Pfaffian coordinates, this amounts to $p = 0$. It is then straightforward to evaluate the residue along $p = 0$ of \eqref{eq:pfaffian3pt} to find
\be
    \mathop{\rm Res}_{p = 0} \omega_{\geq 0} = -\operatorname{dlog}p_{13} \wedge \operatorname{dlog}p_{23} + \operatorname{dlog}p_{12} \wedge \operatorname{dlog}p_{23} - \operatorname{dlog}p_{12} \wedge \operatorname{dlog}p_{13} = -\frac{\omega_{\mathbb{P}^2}}{p_{12} \hs p_{13} \hs p_{23}} \,,
\ee
which is the correct flat-space form after identifying each $p_{ij}$ with the minor $(ij)$ of a matrix representative $\mathcal{C} \in \gr(2,3)$.

\vskip 4pt
Before moving on, we briefly study this warm-up case from the viewpoint of the mixed Hodge theory approach, as it showcases the strategy we are going to pursue later on.
In this case, we take the algebraic boundary $\partial_a\PP^3_{\geq 0}$ of the orthant as the subvariety $D$ in $\PP^3$.
The boundary stratification $\mathcal{L}(\partial_a\PP^3_{\geq 0})$ is simply the face poset of the $3$-simplex, with all irreducible components being a projective space $\PP^k$ for some $k\leq 3$.
Given that $g(\PP^k)=g(\PP^k,\varnothing)=0$ for all $k$, we also have $g(Z,D_Z) = 0$ for any $Z\in \mathcal{L}(D)$.
Moreover, since $\PP^3_{\geq 0}$ is a real region in $\PP^3\!\setminus\!\partial_a\PP^3_{\geq 0}$, with boundaries in $\partial_a\PP^3_{\geq 0}$ by construction, it defines a well-defined cycle $[\PP^3_{\geq 0}]\in H_3(\PP^3,\partial_a\PP^3_{\geq 0})$ with non-zero canonical form.
Hence, $[\PP^3_{\geq0}]$ is a positive geometry.
Note that, since $\PP^3_{\geq 0}$ is a connected regular semi-algebraic set, it also follows from \cite[Proposition 2.4.5]{thesis} that the triple $(\PP^3,\PP^3_{\geq 0},\omega_{\geq 0})$ is a positive geometry as in Definition~\ref{def:rec_PG}. In the next section, we will investigate the case $n=4$ in a similar fashion.

\subsection{Four-Point Function}
The three-point result motivates us to move on to the non-trivial case of four-point correlators. In this case, the investigation of positive geometries in the orthogonal Grassmannian must be complemented with the physical insights of factorization and flat-space limit. Crucially, the former demands the addition of two Mandelstam hypersurfaces that enrich the kinematic space. The resulting structure is the playground from which the Yang--Mills correlators will then arise as positive geometries.

\subsubsection{Four Regions}
Recall that, for $n=4$, the embedding $\varphi:\OGr_R(4,8)\dashrightarrow \PP^7$ in \eqref{eq:fourpointmap} is given by the Pfaffian coordinates $\{ p, \hs p_{ij}, \hs p_{1234} \}$ that satisfy the \emph{Pfaffian constraint}
\begin{align}
    \label{eq:pfaffian_constraint}
    p \hs p_{1234} - p_{12} \hs p_{34} + p_{13} \hs p_{24} - p_{14} \hs p_{23}=0 \,.
\end{align}
As a first step, we again restrict to the subset of $\OGr_R(4,8)$ with non-negative coordinates, $p,\hs p_{ij}, \hs p_{1234} \geq  0$.
In this parametrization of the orthogonal Grassmannian, the Mandelstams take the form
\be
\begin{alignedat}{3}\label{eq:S_and_T}
    S & \equiv(\bar 1 \bar 2 1 2) && = p_{13} \hs p_{24} -p_{14} \hs p_{23} && = p_{12} \hs p_{34} - p \hs p_{1234} 
    \,, \\[4pt]
    T & \equiv(\bar 1 \bar 4 1 4) && = p_{13} \hs p_{24} - p_{12} \hs p_{34} && = p_{14} \hs p_{23} - p \hs p_{1234} \,.
\end{alignedat}
\ee
The hypersurfaces $S = 0$ and $T = 0$, whose addition is motivated by factorization, split the positive orthant into four regions or, more precisely, four semi-algebraic sets. In each region, $S$ and $T$ have a fixed sign.
Put in symbols, we set
\begin{align}
    R^{\pm\pm} \equiv \left\{ (p : p_{ij} : p_{1234}) \in \OGr_R(4,8) \mid p_I\geq 0,\: \pm S(p)\geq 0,\: \pm T(p)\geq 0\:\right\}\subset \OGr_R(4,8) \,.
\end{align}
These four regions are isomorphic to each other, which follows from the symmetries in the defining equations.
For example, we get an isomorphism between $R^{++}$ and $R^{--}$ by interchanging the coordinates 
\be
    p \leftrightarrow p_{13} \quad \text{and} \quad p_{24}\leftrightarrow p_{1234} \,,
\ee
and leaving the other coordinates unchanged. This sends 
\be
    S= p_{13} \hs p_{24} - p_{14} \hs p_{23} \; \mapsto \;  p \hs p_{1234} - p_{14} \hs p_{23} = -T \,,
\ee
and conversely $T\mapsto -S$. The maps for the other regions are analogous. Moreover, these semi-algebraic sets are connected and full-dimensional, as can be computed in \emph{Mathematica}.
Similarly, we compute the algebraic boundaries, the residual arrangement and the canonical form of each region; see Table~\ref{tab:regions_CF_OGR}. Throughout this section, we will frequently use the shorthand
\be
    \label{eq:d6pI}
    \dif^6 p_I \equiv \dif p\wedge \dif p_{1234}\wedge \dif p_{12}\wedge \dif p_{23}\wedge \dif p_{34}\wedge \dif p_{14} \,,
\ee
corresponding to the affine chart with $p_{13}=1$, where $p_{24}$ is fixed by the Pfaffian constraint \eqref{eq:pfaffian_constraint}.

\begin{table}
    \centering
    \renewcommand{\arraystretch}{1.9}
    \begin{tabular}{lccc}
        Region & Algebraic Boundary & Residual Arrangement & Canonical Form\\
        \midrule
        $R^{++}$ 
            & $\{p \hs p_{1234} \hs S\hs T =0\}$ & $\varnothing$
            & $\displaystyle \frac{-1}{p \hs p_{1234} \hs S\hs T}\:\dif^6p_I$ \\
        $R^{+-}$ 
            & $\{p_{14} \hs p_{23} \hs S\hs T =0\}$ & $\varnothing$
            & $\displaystyle \frac{1}{p_{14} \hs p_{23} \hs S \hs T}\:\dif^6p_I$ \\
        $R^{-+}$ 
            & $\{p_{12} \hs p_{34} \hs S\hs T=0\}$ & $\varnothing$
            & $\displaystyle \frac{1}{p_{12} \hs p_{34} \hs S\hs T}\:\dif^6p_I$ \\
        $R^{--}$ 
           & $\{p_{13} \hs p_{24} \hs S\hs T =0\}$ & $\varnothing$
            & $\displaystyle \frac{-1}{p_{13} \hs p_{24} \hs S \hs T}\:\dif^6p_I$
    \end{tabular}
    \caption{Algebraic boundary, residual arrangement, and canonical forms of $R^{\pm\pm}$.}
    \label{tab:regions_CF_OGR}
\end{table}

\vskip 4pt
Given the algebraic boundary, the (putative) canonical forms should not be surprising. The denominator is expected to be the defining equation of $\partial_a R^{\pm \pm}$, as demanded by the need for logarithmic poles, and we recall that the degree of the polynomial in the numerator is fixed to be
\be
    \deg \operatorname{adj}(R^{\pm \pm}) = \deg \partial_aR^{\pm\pm} + d - n-1 \,,
\ee
where in this case $\deg \partial_aR^{\pm \pm} = 6$ is due to the algebraic boundary, $d = 2$ is the degree of the defining equation \eqref{eq:pfaffian_constraint}, and $n = 7$ is the dimension of the ambient space $\PP^7$. Together, these imply $\deg \operatorname{adj}(R^{\pm \pm}) = 0$, i.e.~that $\operatorname{adj}(R^{\pm \pm})$ is a constant.

\vskip 4pt
In the next subsection, we will prove that $[R^{\pm\pm}]\in H_6(\OGr_R(4,8),D)$ are positive geometries with the canonical forms above. Following the examples in the previous section, we begin the analysis by characterizing the boundary stratification of this setup.

\subsubsection{Algebraic Boundary}\label{subsec:alg_bd}
In the mixed Hodge theory approach to positive geometries, characterizing the boundary stratification of $R$ is a necessary step to describe the residues along its faces and verify the recursive vanishing genus requirement. To this end, we start by noting that the union of the algebraic boundaries of the four regions consists of all eight coordinate hyperplane sections in $\OGr_R(4,8)$, together with $S=0$ and $T=0$.
Let $D\subset \OGr_R(4,8)$ be the codimension-$1$ subvariety given by the union of these irreducible components, i.e.
\be
    \label{eq:ourdivisor}
   D = \{p=0\}\cup \{p_{1234}=0\}\cup \ldots \cup\{p_{34}=0\}\cup \{S=0\}\cup \{T=0\}\subset \OGr_R(4,8) \,.
\ee
In order to show that the regions $R^{\pm\pm}$ from the previous subsection are positive geometries, we study the algebraic boundary $D\subset\OGr_R(4,8)$ and its stratification.
Our main result in this section is that the boundary stratification $\mathcal{L}(D)$ has $150$ irreducible components, all of which have vanishing genus, i.e.~$g(Z)=g(Z,\varnothing)=0$ for all $Z\in \mathcal{L}(D)$.
This implies by \cite[Corollary 2.4.3]{thesis} that $g(Z,D_Z)=0$ for all $Z\in\mathcal{L}(D)$, hence $[R^{\pm\pm}]\in H_6(\OGr_R(4,8),D)$ are positive geometries as in Definition \ref{def:MHS_PG}.
In particular, since these regions are regular, connected and orientable semi-algebraic sets, they also give rise to positive geometries as in Definition~\ref{def:rec_PG}; see \cite[Proposition 2.4.5]{thesis}.

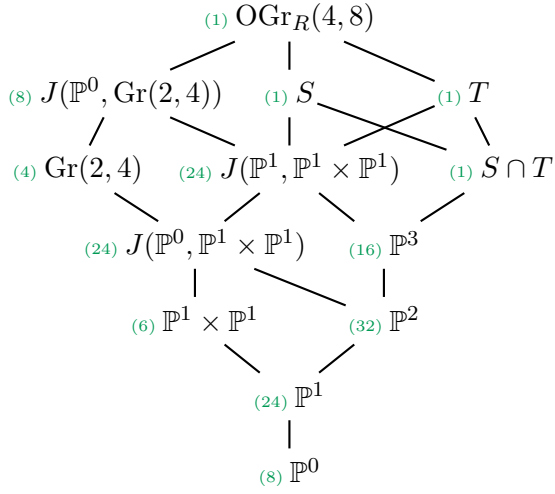
\begin{figure}
  \centering
  \begin{tikzpicture}[
      every node/.style={
        rectangle,
        inner sep=.5pt,
        minimum size=6mm
      },
      every path/.style = {
        thick,
        smooth
      },
      layer/.style={yshift=#1cm}
    ]

    \node (D) at (0,6) {\textcolor{ForestGreen}{\tiny{$(1)$}}\:$\OGr_R(4,8)$};
    
    \node (a1) at (-2.3,5) {\textcolor{ForestGreen}{\tiny{$(8)$}}\:$J(\PP^0,\gr(2,4))$};
    \node (a2) at (0,5) {\textcolor{ForestGreen}{\tiny{$(1)$}}\:$S$};
    \node (a3) at (2.3,5) {\textcolor{ForestGreen}{\tiny{$(1)$}}\:$T$};
    
    \node (b1) at (-2.8,4) {\textcolor{ForestGreen}{\tiny{$(4)$}}\:$\gr(2,4)$};
    \node (b2) at ( 0,4) {\textcolor{ForestGreen}{\tiny{$(24)$}}\:$J(\PP^1, \PP^1\times\PP^1)$};
    \node (b3) at ( 2.8,4) {\textcolor{ForestGreen}{\tiny{$(1)$}}\:$S\cap T$};
    
    \node (c1) at (-1.25,3) {\textcolor{ForestGreen}{\tiny{$(24)$}}\:$J(\PP^0, \PP^1\times\PP^1)$};
    \node (c2) at ( 1.25,3) {\textcolor{ForestGreen}{\tiny{$(16)$}}\:$\PP^3$};
    
    \node (d1) at (-1.25,2) {\textcolor{ForestGreen}{\tiny{$(6)$}}\:$\PP^1\times \PP^1$};
    \node (d2) at (1.25,2) {\textcolor{ForestGreen}{\tiny{$(32)$}}\:$\PP^2$};
    
    \node (e1) at (0,1) {\textcolor{ForestGreen}{\tiny{$(24)$}}\:$\PP^1$};
    
    \node (f1) at (0,0) {\textcolor{ForestGreen}{\tiny{$(8)$}}\:$\PP^0$};
    
    \draw (f1) -- (e1);
    
    \draw (e1) -- (d2);
    \draw (e1) -- (d1);
    
    \draw (d1) -- (c1);
    \draw (d2) -- (c2);
    \draw (d2) -- (c1);
    
    \draw (b1) -- (c1);
    \draw (b2) -- (c1);
    \draw (b2) -- (c2);
    \draw (b3) -- (c2);

    \draw (a1) -- (b1);
    \draw (a1) -- (b2);
    \draw (a2) -- (b2);
    \draw (a2) -- (b3);
    \draw (a3) -- (b3);
    \draw (a3) -- (b2);

    \draw (D) -- (a1);
    \draw (D) -- (a2);
    \draw (D) -- (a3);
    
  \end{tikzpicture}
  \caption{Algebraic boundary stratification of $D$.}
  \label{fig:alg_bd}
\end{figure}

\vskip 4pt
The boundary stratification of $D$ is shown in Figure~\ref{fig:alg_bd}, where the small green numbers indicate the number of strata of that type.
Recall that we obtain the strata in $\mathcal{L}(D)$ by intersecting irreducible components of $D$ with each other.
We now explain in detail how to obtain the boundary stratification. 
This stratification can also be directly computed via a computer algebra system such as \emph{Macaulay2}.

\vskip 4pt
First, recall that the orthogonal Grassmannian $\OGr_R(4,8)$ in its Pfaffian embedding is the quadric hypersurface in $\PP^7$ defined in \eqref{eq:pfaffian_constraint}. 
We can obtain the defining equations of $\gr(2,4)\subset \PP^5$ by setting the two coordinates of any monomial in \eqref{eq:pfaffian_constraint} to zero. For example, setting $p = p_{1234} = 0$, we get
\begin{align}\label{eq:subGrass}
    - p_{12} \hs p_{34} + p_{13} \hs p_{24} - p_{14} \hs p_{23} =0 \,.
\end{align}
In terms of $\gr(2,4)$, each $p_{ij}$ is interpreted as the minor $(ij)$ of the $2 \times 4$ matrix representative $\mathcal{C} \in \gr(2,4)$. Hence, there are $\binom{4}{1}=4$ such strata.
Similarly, we obtain the quadric surface $\PP^1\times \PP^1\subset \PP^3$ by setting all the Pfaffian coordinates of two monomials to zero in \eqref{eq:pfaffian_constraint}, e.g.
\begin{align}\label{eq:subSegre}
    p_{13} \hs p_{24} - p_{14} \hs p_{23}=0 \,.
\end{align}
In this example, we have set $p = p_{1234}=p_{12}=p_{34}=0$, so that $(p_1:p_2) \times (p_3:p_4) \in \mathbb{P}^1 \times \mathbb{P}^1$ maps to $(p_{ij} = p_i p_j) \in \mathbb{P}^3$, with $i=1,2$ and $j=3,4$, i.e.~points in $\PP^3$ that obey the above constraint. Hence, there are $\binom{4}{2}=6$ of these strata.
The projective spaces $\P^3$ are obtained by setting a single coordinate of each monomial in \eqref{eq:pfaffian_constraint} to zero.
Hence there are $2^4=16$ of these strata.
The counting for the other projective spaces is similar.

\vskip 4pt
We now turn to the slightly more subtle strata in Figure~\ref{fig:alg_bd}.
For this, we need the definition of joins of varieties.
Let $X,Y\subset \PP^N$ be two projective varieties.
Their join $J(X,Y)\subset \PP^N$ is the set of all points in $\PP^N$ lying on any line $\overline{xy}\equiv \{a x+b y \mid (a:b)\in \mathbb P^1\}$, with $x\in X$ and $y\in Y$.
Put in symbols, we have
\be
    J(X,Y) \equiv \bigcup_{(x,y)\in X\times Y}\overline{xy}\subset \PP^N \,.
\ee
Our first stratum is then $J(\PP^0,\gr(2,4))\subset \PP^6$, denoting the join of the Grassmannian with a base point $e_I \cong \PP^0$, which corresponds to any of the coordinate vertices in $\PP^7$.
More precisely, $J(\PP^0,\gr(2,4))$ is the union of lines going through one of the coordinate points $e_I$ of $\PP^7$ and a point in the Grassmannian, which is obtained by deleting the monomial in the Pfaffian constraint \eqref{eq:pfaffian_constraint} that contains $e_I$.
Hence, there are $8$ of these strata.
For example, $J(e_{\varnothing},\gr(2,4))=\{p_{1234}=0\}\subset \OGr_R(4,8)$ is the join of $e_{\varnothing}$ with the Grassmannian $\gr(2,4)$ whose defining equation is \eqref{eq:subGrass}.\footnote{To see this explicitly, consider an element $(p:0:p_{ij})$ in $\{ p_{1234} = 0 \}$. In the chart $p = 1$, the overall scale of $p_{ij}$ is not projective, so we can write $p_{ij} = t q_{ij}$, with $\{q_{ij} \} \in \gr(2,4) \subset \mathbb{P}^5$. The line connecting $e_\varnothing$ with $\{q_{ij}\}$ is then parametrized precisely by $t$. We can restore the projective scaling by writing $(s:0:r q_{ij})$, with $(s:r) \in \mathbb{P}^1$ a projective parametrization of the line. This reasoning can be applied to the other joins as well.}

\vskip 4pt
Next, $J(\PP^1,\PP^1\times \PP^1)$ denotes the join of $\PP^1$ with $\PP^1\times\PP^1$, i.e.~it is the set traced out by all the projective $2$-planes containing the base $\PP^1$ and a point on $\PP^1\times \PP^1$.
The base $\PP^1$ is spanned by any pair of the vertices of $\PP^7$, except for the four lines corresponding to the four monomials in the Pfaffian constraint.
Said otherwise, $p \hs p_{1234}$, $p_{12} \hs p_{34}$, $p_{13} \hs p_{24}$, and $p_{14} \hs p_{23}$ do not give rise to such a stratum.
Hence, there are $\binom{8}{2}-4=24$ of these strata. An example is obtained by setting $p = p_{12} = 0$, so that $\PP^1 \times \PP^1$ is realized as the quadric surface \eqref{eq:subSegre}, and $(p_{1234} : p_{34})$ describes the base $\PP^1$.

\vskip 4pt
Similarly, we get a stratum of type $J(\PP^0,\PP^1\times\PP^1)$ for each binomial in the Pfaffian constraint and choice of three vanishing coordinates that are not in the support of the binomial, that is, $\binom{4}{2}\cdot \binom{4}{3}=24$ strata. In the example above for $J(\PP^1,\PP^1\times \PP^1)$, one such stratum can be obtained by further setting $p_{34} = 0$.
The remaining counts for the strata are simpler to derive.
In total, we have $150$ distinct strata, i.e.~$|\mathcal{L}(D)|=150$.

\vskip 4pt
Note that all strata appearing in the boundary stratification $\mathcal{L}(D)$ are either projective spaces, Grassmannians, quadrics $\PP^1\times \PP^1$, joins of them, or complete intersections, i.e.~varieties whose codimension matches the number of defining equations, in $\OGr_R(4,8)$.
Therefore, all strata have vanishing genus, that is, $g(Z)=g(Z,\varnothing) = 0$ for all $Z\in\mathcal{L}(D)$; see e.g.~\cite[Chapter 2]{thesis}.
By \cite[Corollary 2.4.3]{thesis}, it then follows that $g(Z,D_Z)=0$ for all $Z\in\mathcal{L}(D)$.
Recall that we computed the algebraic boundary and residual arrangement of $R^{\pm\pm}$ in Table~\ref{tab:regions_CF_OGR}.
We therefore have that $[R^{\pm\pm}]\in H_6(\OGr_R(4,8),D)$ are positive geometries with canonical forms $\omega_{\pm\pm}=\Omega([R^{\pm\pm}])$ as in Table~\ref{tab:regions_CF_OGR}.
Moreover, since the regions $R^{\pm\pm}$ are regular, connected and orientable, by \cite[Proposition 2.4.5]{thesis} they also give rise to a positive geometry triple $(\OGr_R(4,8),R^{\pm\pm},\omega_{\pm\pm})$ as in Definition~\ref{def:rec_PG}.

\vskip 4pt
Before we move on, it is very useful for the residue computations in the following subsection to identify the subvarieties defined by the vanishing of $S$ and $T$ and their intersection, which is in turn defined by their simultaneous vanishing. The discussion that follows is a rigorous implementation of the factorization rules \eqref{equ:minors-factorized} and \eqref{eq:simplegluing} in terms of the Pfaffian coordinates.
We will show that $\{S=0\}\subset\OGr_R(4,8)$ is isomorphic to the quotient $(\PP^3\times \PP^3)//\mathbb{C}^*$\footnote{The double slash indicates that we only have a $\CC^*$-bundle over some dense set in $\PP^3\times \PP^3$, as shown in more detail in \cite[Corollary 6.1.5]{thesis}} with a group action given by \emph{little group transformations}:
\begin{align}\label{eq:torus_action}
  t\bullet [p^L]\times [p^R] = (p^L :p^L_{12}:tp^L_{1s}:tp^L_{2s})\times (p^R:p^R_{34}:t^{-1}p^R_{3s}:t^{-1}p^R_{4s}) \,.
\end{align}
Given that $\PP^3 \cong \OGr_R(3,6)$, this isomorphism is the geometric version of the statement that four-point kinematics factorize into the product of three-point kinematics on the vanishing locus of a Mandelstam. We therefore label the coordinates in a way that reflects the factorization into a left and a right piece, i.e.~$A_4 \sim (A_{3,L} A_{3,R})/S$ as in \eqref{eq:factorule}. The parameter $t$ that we quotient by, in particular, can be interpreted as the action of little group transformations on the exchanged particle in a four-point correlator.
Then, by \cite[Lemma 6.1.6]{thesis}, the quotient $(\PP^3\times \PP^3)//\mathbb{C}^*$ can be identified with 
\begin{align}\label{eq:GIT}
    \{ p_{12} \hs p_{34} - p \hs p_{1234} = p_{13} \hs p_{24} - p_{14} \hs p_{23} =0 \} = \{S=0\}\subset \OGr_R(4,8) \,.
\end{align}
In order to see this, consider the map $\mathcal{S}:\PP^3\times \PP^3\to \P^{15}$ sending
\be
    (p^L:p^L_{12}:p^L_{1s}:p^L_{2s})\times (p^R:p^R_{34}:p^R_{3s}:p^R_{4s}) \; \mapsto \; (p^L \hs p^R : p^L \hs p^R_{34}:\ldots:p^L_{2s} \hs p^R_{4s}) \,,
\ee
which mathematically is a \emph{Segre embedding}.
One can also think of this map as the tensor map $(x,y) \mapsto x \otimes y$.
Denote the coordinates of the codomain $\P^{15}$ by $p_I=p^L_{\alpha}p^R_{\beta}$, with $\alpha\subset \{1,2,s\}$ and $\beta \subset \{3,4,s\}$ of even size, and $I=(\alpha \cup \beta)\!\setminus\! \{s\}$.
For example, we have $p_1 = p_{1s}^L \hs p^R$, $p_{13}=p_{1s}^L \hs p_{3s}^R$ and $p_{134} = p_{1s}^L \hs p_{34}^R$.
Let $\pi:\PP^{15}\to \PP^7$ then be the projection onto the coordinates with variables labeled by all even-sized subsets $I\subset \{1,2,3,4\}$, which only keeps the components that are invariant under the action of the little group transformation~\eqref{eq:torus_action}.
It follows that the Zariski closure $Y\subset \PP^7$ of $\operatorname{im}\varphi$, with 
\begin{align}\label{eq:PP^3_to_S}
    \varphi \equiv\pi \circ \mathcal{S}: \PP^3\times \PP^3 \longrightarrow \PP^7 \,,
\end{align}
is the five-dimensional variety cut out by the determinantal Segre relations on the coordinates of $\PP^7$, i.e.~we have
\begin{align}\label{eq:Y_cosmology}
  \overline{\operatorname{im}\varphi}
  =\{ p_{12} \hs p_{34} - p \hs p_{1234} = p_{13} \hs p_{24} - p_{14} \hs p_{23} =0 \}\subset \PP^7 \,,
\end{align}
which are the defining equations of $\{S=0\}\subset \OGr_R(4,8)$, as in \eqref{eq:GIT}.

\vskip 4pt
We have just shown that $\{S=0\}\subset \OGr_R(4,8)$ is an embedding of $(\PP^3\times \PP^3)//\mathbb{C}^*$, hence we get an isomorphism $\widetilde \varphi:(\PP^3\times \PP^3)//\mathbb{C}^*\to \{S=0\}\subset \OGr_R(4,8)$ between these two varieties.
The map $\widetilde \varphi$ also restricts to an isomorphism of semi-algebraic sets between $(\PP^3_{\geq 0}\times \PP^3_{\geq 0})//(\RR_{>0})$ and $(R^{++}\cup R^{--})\cap \{S=0\}$. This observation will be crucial in the next subsection. The case of $T$ follows analogously by symmetry in the defining equations of $\OGr_R(4,8)$ and $\{T=0\}$.

\subsubsection{Reduced Correlator}
The previous discussion was instrumental in proving the key point that $R^{\pm\pm}$ are all positive geometries, and that any of their unions is as well. We can now finally identify the region $R$ whose canonical form satisfies the physical requirements of having a flat-space limit and factorizing into the wedge product of $3$-forms on $S = 0$ and $T = 0$, quotiented by the action of little group transformations \eqref{eq:torus_action}.

\vskip 4pt
We first observe that, in the language of minors, the flat-space Grassmannian $\gr(2,4)$ lies on the locus $(\bar1\bar2\bar3\bar4)=(1234) = 0$; see~\cite{Arundine:2026fbr}.
In terms of the Pfaffian coordinates, this translates to $p = p_{1234} = 0$. As shown in Table~\ref{tab:regions_CF_OGR}, these hypersurfaces define two of the boundaries of $R^{++}$. For this reason, we start by considering the region $R^{++}$ and its canonical form $\omega_{++}$, which is also given in Table~\ref{tab:regions_CF_OGR} as
\be
    \omega_{++} = \frac{-1}{p \hs p_{1234} \hs S \hs T} \hs \dif^6p_I \,.
\ee
We now inspect its factorization properties by computing its residue along $S = 0$, which yields
\be\label{eq:R^++_bad}
    \mathop{\rm Res}_{S=0} \omega_{++} = \frac{-1}{p \hs p_{1234} \hs p_{34} \hs T} \hs \dif^5p_{\widehat{12}} \,,
\ee
where $\dif^5p_{\widehat{12}}$ is the $5$-form obtained from $\dif^6p_I$ in \eqref{eq:d6pI} by omitting $\dif p_{12}$. Because of the presence of $T$ in the denominator, we conclude that this form is incompatible with the required factorization property, as this cannot be expressed as the wedge product of two $3$-forms $\omega_3$, quotiented by the action of little group transformations. Indeed, in the coordinates \eqref{eq:torus_action} of the lower-point $\OGr_R(3,6)$, we have
\be
T|_{S=0} = (p_{1s}^L \hs p_{4s}^R)(p_{2s}^L \hs p_{3s}^R) - (p^L \hs p^R) (p_{12}^L \hs p_{34}^R) \,,
\ee
which is an irreducible polynomial, in particular not of the form $f\big(p^L\big) \hs g\big(p^R\big)$ and hence an obstruction to factorization. 

\vskip 4pt
These observations teach us that $R^{++}$ is a necessary element of the positive region $R$, but it is only part of it. We have seen in the toy example of the two simplices from Figure~\ref{fig:twoSimplices} that we can sometimes join two regions along a shared facet to get rid of spurious poles. It is therefore natural to consider the region $R=R^{++}\cup R^{--}$, which is also the \textit{only} union of $R^{++}$ with another region whose canonical form preserves both the $S = 0$ and $T = 0$ boundaries.
This operation returns
\be\label{eq:CF_w_R}
    \boxed{\omega_R = \Omega([R^{++}]- [R^{--}])=\frac{-(S+T)}{p \hs p_{1234} \hs p_{13} \hs p_{24} \hs S \hs T} \hs \dif^6p_I} \;.
\ee
The relative sign is necessary to cancel the residue on $S = T = 0$, which poses an obstruction to factorization, as was already explained for the origin in the toy example.
Indeed, the numerator of $\omega_R$ cancels precisely all the spurious poles associated with $R=R^{++}\cup R^{--}$. More specifically, it is the unique quadratic form in the Pfaffian coordinates that interpolates the residual arrangement
\be
    \mathcal{R}(R) = \{p=p_{13}=0\}\cup \{p=p_{24}=0\}\cup \{p_{1234}=p_{13}=0\}\cup \{p_{1234}=p_{24}=0\} \,,
\ee
and $\{S=T=0\}$.

\vskip 4pt
The physical requirements we have imposed have unambiguously identified $\omega_R = \hat A_4 \hs \dif^6 p_I$, with $\hat A_4$ the helicity-stripped reduced Yang--Mills correlator in \eqref{eq:YMredref}!
It is therefore expected that both the factorization property and the flat-space limit hold. Indeed, we can take the residue along $S=0$ to obtain
\be
    \label{eq:resS}
    \mathop{\rm Res}_{S=0} \omega_R = \frac{-1}{p \hs p_{1234} \hs p_{13} \hs p_{24} \hs p_{34}} \hs \dif^5p_{\widehat{12}} \,.
\ee
This residue is the fiber integration $\varphi_*$ of the canonical form $\omega^L_{\geq 0}\wedge\omega^R_{\geq 0}$ of $\PP^3_{\geq 0}\times \PP^3_{\geq 0}$ from \eqref{eq:pfaffian3pt} along the one-dimensional fibers of $\varphi$ from \eqref{eq:PP^3_to_S}, parametrized by the little group parameter $t\in \CC^*$ defined in \eqref{eq:torus_action}:
\be
    \mathop{\rm Res}_{S=0} \omega_R 
    = \varphi_*\bigg(\frac{\omega^L_{\PP^3}}{p^L \hs p^L_{12} \hs p^L_{1s} \hs p^L_{2s}} \wedge \frac{\omega^R_{\PP^3}}{p^R \hs p^R_{34} \hs p^R_{3s} \hs p^R_{4s}} \bigg) 
    = \frac{-1}{p \hs p_{1234} \hs p_{13} \hs p_{24} \hs p_{34}} \hs \dif^5p_{\widehat{12}} \,.
\ee
This fiber integration of the form is well-defined, since we can write the differential form on the orthants in terms of the invariant coordinates as follows:
\begin{align}
    \omega^L_{\geq 0}\wedge\omega^R_{\geq 0} & = \big(\operatorname{dlog}p^L \wedge \operatorname{dlog}p_{12}^L \wedge \operatorname{dlog}p_{2s}^L \big) \wedge \big(\operatorname{dlog} p^R \wedge \operatorname{dlog} p_{34}^R \wedge \operatorname{dlog} p_{4s}^R\big) \nonumber \\[4pt]
    & = -\operatorname{dlog}t \wedge \operatorname{dlog}p\wedge \operatorname{dlog}p_{1234}\wedge \operatorname{dlog}p_{23}\wedge \operatorname{dlog}p_{34}\wedge \operatorname{dlog}p_{14} \nonumber \\[4pt]
    & = \operatorname{dlog} t \wedge \frac{-1}{p \hs p_{1234} \hs p_{14} \hs p_{23} \hs p_{34}} \hs \dif^5p_{\widehat{12}}  = \operatorname{dlog} t \wedge \frac{-1}{p \hs p_{1234} \hs p_{13} \hs p_{24} \hs p_{34}} \hs \dif^5p_{\widehat{12}} \,.
\end{align}
In the above calculation, we have worked in the chart $p_{1s}^L = p_{3s}^R = 1$ and have set $p^L = t^{-1}$, which follows from the projective transformation that restores $p_{1s}^L = 1$ after acting with \eqref{eq:torus_action}. The final step uses the defining equations \eqref{eq:Y_cosmology} of $\{S=0\}$. Next, recall that the image of $\varphi$ is the five-dimensional locus $\{S = 0\}$ in $\OGr_R(4,8)$, which projects out the little group parameter $t$.
We can therefore write
\be
    \label{eq:formfact}
    \mathop{\rm Res}_{S=0} \omega_R = \varphi_*\big(\omega^L_{\geq 0} \wedge \omega^R_{\geq 0} \big)\,,
\ee
with $\omega^{L,R}_{\geq 0}$ the three-point form \eqref{eq:pfaffian3pt}. This equation realizes precisely the factorization condition \eqref{eq:factorule} in terms of forms in the Pfaffian embedding of the orthogonal Grassmannian.

\vskip 4pt
In \eqref{eq:Y_cosmology}, we established the isomorphism $\widetilde\varphi$ between $(\PP^3\times\PP^3)//\CC^*$ and $\{S=0\}\subset \OGr_R(4,8)$. Moreover, it restricts to an isomorphism of the semi-algebraic sets $(\PP^3_{\geq 0}\times\PP^3_{\geq 0})//\RR_{>0}$ and $(R^{++}\cup R^{--})\cap \{S=0\}$.
As a consequence, taking residues of the canonical form $\omega^L_{\geq 0}\wedge \omega^R_{\geq 0}$ along the boundaries of $\PP^3_{\geq 0}\times\PP^3_{\geq 0}$ commutes with applying the map $\varphi$. The factorization on $T = 0$ is also satisfied.

\vskip 4pt
We now check that we correctly recover the Yang--Mills scattering amplitude on $p = p_{1234} = 0$. First, consider the residue along $p_{1234} = 0$.
Geometrically, the set $\{p_{1234}=0\}$ is the join $J(e_\varnothing,\gr(2,4))$ of the Grassmannian
\be
    \{- p_{12}p_{34} + p_{13}p_{24} - p_{14}p_{23} = 0\}\subset \PP^5 \,,
\ee
with coordinates $\{ p_{12},p_{13},p_{14},p_{23},p_{24},p_{34} \}$, with a base point $e_\varnothing \in \PP^7$; see also Figure~\ref{fig:alg_bd}.
Its intersection with $R$ yields $J(e_{\varnothing},\gr(2,4)_{\geq 0})$, the join of the non-negative Grassmannian $\gr(2,4)_{\geq0}$ with $\{e_{\varnothing}\}\cong \PP^0$.
Given that $S + T = p_{13} \hs p_{24} - p \hs p_{1234}$ on $\OGr_R(4,8) \subset \PP^7$, we then obtain
\begin{align}
    \mathop{\rm Res}_{p_{1234} = 0} \omega_R 
    &= \frac{1}{p}\dif p\wedge \frac{p_{13} \hs p_{24} - p \hs p_{1234}}{p_{13} \hs p_{24} \hs S \hs T}\bigg\vert_{p_{1234}=0}  \hs \dif p_{12}\wedge \dif p_{23}\wedge \dif p_{34}\wedge \dif p_{14} \nonumber\\[4pt]
    &= \frac{1}{p}\dif p\wedge\frac{1}{S \hs T}\bigg\vert_{p_{1234}=0} \hs \dif p_{12}\wedge \dif p_{23}\wedge \dif p_{34}\wedge \dif p_{14} \nonumber\\[4pt]
    &= \frac{1}{p}\dif p \wedge \frac{1}{p_{12} \hs p_{23} \hs p_{34} \hs p_{14}} \hs \dif p_{12}\wedge \dif p_{23}\wedge \dif p_{34}\wedge \dif p_{14} \,,
\end{align}
where the sign arises from placing $\dif p_{1234}$ as the first element in the wedge product before taking the residue, as prescribed in \eqref{eq:residuemap}. This is clearly the canonical form of $J(e_{\varnothing},\gr(2,4)_{\geq 0})$, i.e.~$\operatorname{dlog} p \wedge \omega_{\gr(2,4)}$, after identifying each Pfaffian coordinate $p_{ij}$ as the associated Plücker coordinate $(ij)$ in $\gr(2,4)$. Naturally, we also obtain the canonical form of the non-negative Grassmannian $\gr(2,4)_{\geq 0}$ by taking the residue along $p = 0$: 
\be
    \mathop{\rm Res}_{p = 0} \mathop{\rm Res}_{p_{1234} = 0} \omega_R = \frac{1}{p_{12} \hs p_{23} \hs p_{34} \hs p_{14}} \hs \dif p_{12}\wedge \dif p_{23}\wedge \dif p_{34}\wedge \dif p_{14} \,.
\ee
This result, which is precisely the scattering amplitude demanded by the flat-space limit, does not change by swapping the order of residues (up to the expected orientation sign).

\vskip 4pt
Summarizing, we have found the unique region $R = R^{++} \cup R^{--}$ in $\OGr_R(4,8)$ that satisfies factorization and the flat-space limit. This is a positive geometry, and the canonical function is the (helicity-stripped) reduced Yang--Mills four-point function.
Note that the mixed Hodge theory framework tells us that $[R]\in H_6(\OGr_R(4,8),D)$ is a positive geometry without performing any residue computation. Still, we can compute all the missing iterated residues to also show that $R$ gives rise to a positive geometry $(\OGr_R(4,8),R,\omega_R)$ as in Definition \ref{def:rec_PG}, which we do below. 

\vskip 4pt
In codimension 1, $R$ has two distinct types of boundaries, i.e.~the Mandelstam boundaries and the Pfaffian coordinate boundaries. We have already computed the residue on $S = 0$. Working in the product space $\PP^3_{\geq 0}\times \PP^3_{\geq 0}$, as mentioned earlier, allows us to easily conclude that the triple $(\OGr_R(4,8) \cap \{ S = 0 \},R\vert_{S=0},\operatorname{Res}_S(\omega_R))$ is a positive geometry as in Definition \ref{def:rec_PG}.
Analogous considerations apply to $\{T = 0\}$.
Similarly, we have shown that $\{ p_{1234} = 0\} \cap R \cong J(e_{\varnothing},\gr(2,4)_{\geq 0})$, and that the residue computation returns the associated canonical form.
It is then easy to verify that the same applies to the cases $p=0$, $p_{13}=0$ and $p_{24}=0$. Since each codimension-$1$ boundary of $R$ is a positive geometry, we conclude that the triple $(\OGr_R(4,8),R,\omega_R)$ itself is a positive geometry as in Definition \ref{def:rec_PG}, as had to be shown. Indeed, note that the uniqueness of the form follows from the discussion in Section \ref{subsec:alg_bd}.

\subsubsection{Color-Ordered Correlator}
In the previous subsection, we have characterized the helicity-stripped reduced correlator \eqref{eq:YMredref} as the unique physical form that can arise from the union of regions. We now want to show that the correlator~\eqref{eq:fullcolor} also has a geometric interpretation in the mixed Hodge theory framework. Just like its reduced counterpart, it satisfies the flat-space limit and the factorization condition. However, it is also subject to the \textit{Kleiss-Kuijf relation}~\cite{Kleiss:1988ne}
\be
    \label{eq:KKrel}
    A_4(S,T,U) + A_4(T,U,S) + A_4(U,S,T) = 0 \,,
\ee
with $U = (\bar1\bar3 1 3) = -p_{14}\hs p_{23}-p_{12} \hs p_{34}$. It is easy to show that the form \eqref{eq:CF_w_R} does not satisfy this relation, owing to the identities
\be
\label{eq:STUident}
\begin{alignedat}{3}
    2p \hs p_{1234} & = -S-T-U \,, \qquad && 2p_{12} \hs p_{34} && = S-T-U \,, \\[4pt]
    2p_{13} \hs p_{24} & = S+T-U \,, \qquad && 2p_{14} \hs p_{23} && =-S+T-U \,.
\end{alignedat}
\ee
Indeed, we obtain \eqref{eq:CF_w_R} as the unique answer if we only allow for \textit{unions of regions}. Recall, however, that Definition~\ref{def:MHS_PG} identifies \textit{any} cycle in the homology group $H_n(X(\mathbb{R}),D(\mathbb{R}))$ with a canonical form as a positive geometry. Given its structure as a vector space, then, it is extremely restrictive to only consider unions of regions, as \emph{any} formal linear combination of the regions belongs to the homology group.

\vskip 4pt
Given this observation, we now want to identify the cycle $[\sigma_{KK}] = \sum_{\pm\pm} c_{\pm \pm}[R^{\pm \pm}]$, with canonical form
\be
    \Omega([\sigma_{KK}]) = \sum c_{\pm \pm} \hs \Omega([R^{\pm \pm}]) \,,
\ee
that is also compatible with the condition \eqref{eq:KKrel}. Using the identities \eqref{eq:STUident} and the forms in Table~\ref{tab:regions_CF_OGR}, the form is given by
\be
    \Omega([\sigma_{KK}]) = \frac{2}{S \hs T} \bigg( \frac{c_{++}}{S+T+U} + \frac{-c_{--}}{S+T-U} + \frac{-c_{-+}}{T+U-S} + \frac{-c_{+-}}{U+S-T} \bigg) \hs \dif^6 p_I \,.
\ee
Imposing the validity of \eqref{eq:KKrel} yields the following relations among the coefficients:
\be
    c_{++} = -3 c_{--}, \quad c_{-+} = c_{+-} = -c_{--}\,.
\ee
Interestingly, any non-trivial solution automatically encodes the flat-space limit, as $c_{++} \neq 0$ necessarily, and has the required factorization structure \eqref{eq:formfact} up to normalization. The correct normalization finally fixes $c_{--} = -1/2$. This choice of coefficients correctly reproduces \eqref{eq:fullcolor}, and characterizes it as the canonical function
\be
    [\sigma_{KK}] = \frac{3}{2}[R^{++}] - \frac{1}{2}[R^{--}] + \frac{1}{2}[R^{-+}] + \frac{1}{2}[R^{+-}]\,.
\ee 
As a final remark, we observe that this cycle, hence its canonical form, can be rearranged as follows:
\be
    [\sigma_{KK}] = \frac{1}{2}\big( [R^{++}] - [R^{--}] \big) + \frac{1}{2} \big( [R^{++}] + [R^{+-}] \big) + \frac{1}{2} \big( [R^{++}] + [R^{-+}] \big) \,,
\ee
whose form corresponds to the oriented sum of the three possible reduced correlators, one for each Mandelstam. Up to the factor of $1/2$, each term in the sum also identifies a positive geometry as in Definition~\ref{def:rec_PG}.

\vskip 4pt
This result exemplifies how the broader range of applicability of Definition~\ref{def:MHS_PG}, as opposed to Definition~\ref{def:rec_PG}, might be necessary for the correct interpretation of future physical results. The original definition, in fact, would fail to capture the geometric structure behind the full color-ordered Yang--Mills correlator. It would therefore be intriguing to investigate the consequences of these observations for higher-point functions.

\subsection{Little Group-Invariant Projection}
\label{sec:hyperplane_model}
In the previous subsections, we have studied positive regions in $\OGr_R(n,2n)$, thus discovering the geometric meaning behind Yang--Mills correlators for $n = 3$ and $n = 4$. The characterization of boundary stratifications and the computation of residues has been carried out effectively in terms of the Pfaffian coordinates. There exists, however, a simpler, more intuitive picture that can be used to compute quantities that only depend on the properties of the open subset $X \!\setminus\! D$.

\vskip 4pt
In this subsection, we leverage the little group transformations to establish a model for the four-point geometries discussed previously which simplifies some of the computations and allows us to determine the combinatorial rank.

\subsubsection{Hyperplane Model}
It is possible to describe the open part of the orthogonal Grassmannian $\OGr_R(4,8)\!\setminus\! D$ in terms of a hyperplane arrangement in the plane. To this end, the torus, or little group, action will play a crucial role.

\vskip 4pt
The four-dimensional algebraic torus $(\mathbb{C}^*)^4$ naturally acts on $\OGr(4,8)$ by scaling the columns of a $4 \times 8$ matrix representative. Concretely, we consider
\be
    (\mathbb{C}^*)^4 \cong \left\{{\rm diag}(t_1^{-1},t_2^{-1},t_3^{-1},t_4^{-1},t_1,t_2,t_3,t_4): \; t_1,\ldots,t_4 \in \mathbb{C}^{*} \right\} \subset \mathbb{C}^{8\times 8} \,.
\ee
The action of the torus is then simply obtained by matrix multiplication from the right. This is precisely the little group transformation \eqref{eq:LGactionC}.
In Pfaffian coordinates, we can represent it explicitly as
\begin{equation}\label{eq: torus action}
    (t_1,t_2,t_3,t_4) \bullet (p:p_{12}:\cdots : p_{34}: p_{1234}) = (p:t_1t_2p_{12}:\cdots: t_3t_4p_{34}: t_1t_2t_3t_4 p_{1234}) \,.
\end{equation}
Recall that the three Mandelstam variables in terms of the Pfaffian coordinates are
\be
    S= p_{13} \hs p_{24}-p_{14} \hs p_{23}, \quad T = p_{13} \hs p_{24}-p_{12} \hs p_{34}, \quad U = -p_{14} \hs p_{23} - p_{12} \hs p_{34} \,.
\ee
Note that these are projectively invariant under the torus action \eqref{eq: torus action}.
In particular, the map 
\be
    \varphi:\OGr_R(4,8) \to \PP^2,\quad (p:p_{12}:\cdots: p_{34} : p_{1234}) \mapsto (S:T:U)
\ee
sends $(\mathbb{C}^*)^4$-orbits in $\OGr_R(4,8)$, i.e.~sets of points that can be mapped onto each other by a transformation \eqref{eq: torus action}, to a single point in $\PP^2$.

\begin{figure}
    \centering
    \includegraphics[width=0.62\linewidth]{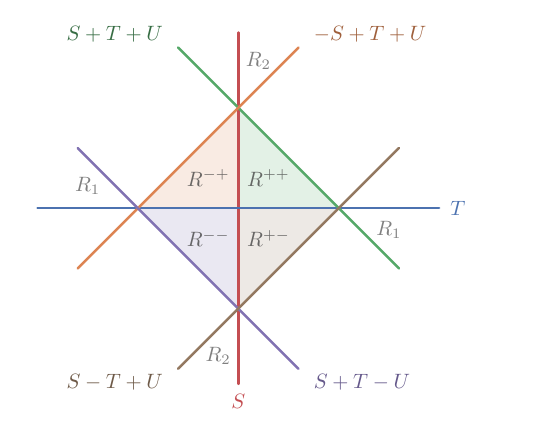}
    \caption{Line arrangement for $\OGr_R^\circ(4,8)/(\mathbb{C}^*)^4$ in the $U=-1$ slice. Each labeled region is bounded by three lines, with $R_{1}$ and $R_2$ being identified projectively.}
    \label{fig: line arrangement}
\end{figure}

\vskip 4pt
We now consider the open part $\OGr_R^\circ(4,8)\equiv\OGr_R(4,8) \!\setminus\! D$, where $D$ is the divisor \eqref{eq:ourdivisor} consisting of all Pfaffian coordinate hyperplanes, together with the vanishing loci of $S$ and $T$. 
The image of $\OGr_R^\circ(4,8)$ under $\varphi$ is the complement of the projective line arrangement
\begin{equation}\label{eq: proj line arrangement}
    \mathcal{A} = \{S, \hs T, \hs S+T+U, \hs -S+T+U, \hs S-T+U, \hs S+T-U \} \,.
\end{equation}
The six lines in $\mathcal{A}$ correspond to $S=0$, $T=0$, and to unions of the hyperplane divisors in $D$ by the identities \eqref{eq:STUident}. Setting $U$ to a fixed constant, for example, determines the global projective scaling. In the slice $U=-1$, the line arrangement $\mathcal{A}$ is shown in Figure~\ref{fig: line arrangement}. This figure exhibits the symmetries of our setup. However, for the purpose of computing the canonical forms of the regions we are interested in, we will use a different chart. A key point is that we have
\begin{align}\label{eq:line_arrangement_isomorphism}
    \OGr_R^\circ(4,8) \cong (\mathbb{C}^*)^4 \times \PP^2\!\setminus\! \mathcal{A} \,. 
\end{align}
Choosing another chart allows us to exhibit the isomorphism $\phi:\OGr_R^\circ(4,8)\rightarrow (\mathbb{C}^*)^4\times \PP^2\!\setminus\! \mathcal{A}$ as
\begin{equation}\label{eq: iso with line arr}
\begin{gathered}
\left( \frac{-S-T-U}{2p_{1234}}: p_{12}: \frac{S+T-U}{2p_{24}}: p_{14}: \frac{-S+T-U}{2p_{14}}: p_{24}: \frac{S-T-U}{2p_{12}}: p_{1234} \right)
\\[0.5em]
\updownarrow
\\[0.5em]
(p_{12},p_{14},p_{24},p_{1234})\times(S:T:U) \,.
\end{gathered}
\end{equation}

\noindent
As written, the upward arrow in \eqref{eq: iso with line arr} is not a well-defined map on $(\mathbb{C}^*)^4 \times \PP^2$, as the scaling of the $(S:T:U)$ part is not preserved.
It is important that we are restricted to the complement of the hyperplane arrangement $\mathcal{A}$.
In fact, we can take one of the lines in $\mathcal{A}$ as the line at infinity and restrict ourselves to an affine chart $\CC^2$, where we then remove the remaining five lines.
These are also important for \eqref{eq: iso with line arr} to be an isomorphism, as they guarantee $p_{12}, \hs p_{14}, \hs p_{24}, \hs p_{1234} \neq 0$.
For example, for a fixed constant $c \in \mathbb{C}^*$, the condition $S+T-U=2c$ fixes the projective scaling, making the map well-defined.
To compute the forms of our regions $R^{\pm\pm}$ of Figure~\ref{fig: line arrangement} in the affine chart $S+T-U=2c$, we use the projective form \eqref{eq:projform} and replace $U$ accordingly.

\vskip 4pt
We explicitly do the computation for $R^{++}$ as an example.
Choosing the orientation on $\PP^2$ that corresponds to the counter-clockwise orientation in Figure~\ref{fig: line arrangement}, this yields
\be
    -\frac{U \hs \dif S \wedge \dif T + T \hs \dif U\wedge \dif S + S \hs \dif T\wedge \dif U}{S \hs T\hs (S+T+U)} = \frac{(S+T-U)\hs \dif S\wedge \dif T}{S\hs T (S+T+U)} =  \frac{2c \: \dif S\wedge \dif T}{S \hs T(S+T+U)}\,, 
\ee
where we have set $S+T-U=2c$ to fix the projective scaling, and thus $\dif U = \dif S + \dif T$.
Similar computations yield the second column of Table~\ref{tab:region-forms}.

\vskip 4pt
At the level of differential forms, we can pullback forms on $(\mathbb{C}^*)^4 \times \PP^2\!\setminus\! \mathcal{A}$ via $\phi^*$ to forms on $\OGr_R(4,8)$. We work with the affine chart of $\OGr_R(4,8)$ given by $p_{13}=1$, which corresponds to the $6$-form $\dif^6 p_I = \dif p \wedge \dif p_{1234} \wedge \dif p_{12} \wedge \dif p_{23} \wedge \dif p_{34} \wedge \dif p_{14}$. For ease of computation, we also want to choose an affine chart of $\PP^2$ that naturally maps under \eqref{eq: iso with line arr} to $p_{13}=1$. Hence, we consider $S+T-U=2p_{24}$. 
Now, recall that we associate canonical forms to semi-algebraic regions in $\OGr_R(4,8)$. The $(\RR^*)^4$ action on the real locus does not preserve these regions, where all Pfaffian coordinates are positive, hence we consider only the action of $(\RR_{>0})^4$.
Therefore, when we consider the pullback of a form on $(\mathbb{C}^*)^4 \times \PP^2\!\setminus\! \mathcal{A}$, we take the form ${\rm dlog} \, p_{1234} \wedge {\rm dlog} \, p_{12}  \wedge {\rm dlog} \, p_{14} \wedge {\rm dlog} \, p_{24}$ for the $(\mathbb{C}^*)^4$ factor, that is, the canonical form of $(\mathbb{R}_{> 0})^4$. For the four regions in Figure~\ref{fig: line arrangement} which are relevant for us, we obtain the results in Table~\ref{tab:region-forms}.

\begin{table}
    \centering
    \renewcommand{\arraystretch}{1.9}
    \begin{tabular}{lcc}
        Region & Form on $\PP^2\!\setminus\!\mathcal{A}$ & Form on $\OGr_R^\circ(4,8)$ \\
        \midrule
        $R^{++}$ 
            & $\displaystyle \frac{S+T-U}{S\hs T(S+T+U)}\hs\dif S \wedge \dif T$
            & $\displaystyle \frac{-1}{p \hs p_{1234} \hs S\hs T} \hs \dif^6p_I$ \\
        $R^{+-}$ 
            & $\displaystyle \frac{-(S+T-U)}{S\hs T (S-T+U)}\hs\dif S \wedge \dif T$
            & $\displaystyle \frac{1}{p_{14} \hs p_{23} \hs S \hs T} \hs \dif^6p_I$ \\
        $R^{-+}$ 
            & $\displaystyle \frac{-(S+T-U)}{S \hs T(-S+T+U)}\hs\dif S \wedge \dif T$
            & $\displaystyle \frac{1}{p_{12} \hs p_{34} \hs S\hs T} \hs \dif^6p_I$ \\
        $R^{--}$ 
            & $\displaystyle \frac{-(S+T-U)}{S \hs T(S+T-U)}\hs\dif S \wedge \dif T$
            & $\displaystyle \frac{-1}{p_{13} \hs p_{24} \hs S \hs T}\hs\dif^6p_I$
    \end{tabular}
    \caption{Canonical forms of the four regions $R^{\pm\pm}$ in both the hyperplane model and in $\OGr_R(4,8)$. The forms in the third column, obtained from the forms in the second column via \eqref{eq: iso with line arr}, match Table~\ref{tab:regions_CF_OGR}.}
    \label{tab:region-forms}
\end{table}

\subsubsection{Combinatorial Rank}
The hyperplane model allows us to easily compute the combinatorial rank $\operatorname{cr}(\OGr_R(4,8),D)$, as defined in \eqref{eq:def_cr}.
Recall that $\OGr_R^\circ(4,8)\cong (\CC^*)^4\times \PP^2\!\setminus\!\mathcal{A}$.
In this case, we have
\be
    \operatorname{cr}(\OGr_R(4,8),D) = \operatorname{cr}(\PP^2,\mathcal{A}) \,
\ee
by \cite[Proposition 2.2.19]{thesis}.
The latter is equal to the number of bounded regions of $\mathcal{A}$ with respect to a generic affine chart; see e.g.~\cite[Proposition 6.3]{brown2025positivegeometriescanonicalforms}.
Note that, on such a chart, the line arrangement will look slightly different from what is shown in Figure~\ref{fig: line arrangement}, because of the two pairs of parallel lines in the latter. In fact, by \cite{Zaslavsky}, the number of bounded regions $b$ can be counted as
\be
    b = 1 -\ell +\sum_{p}(m_p-1) \,,
\ee
where $\ell$ is the number of lines and $m_p$ is the number of lines meeting in a point $p\in\mathcal{A}$.
In our case, we have $\ell = 6$ lines, four points with $m_p=3$, and the remaining three intersection points, including the two at infinity, have $m_p=2$.
The number of bounded regions is therefore $b=6$, and the combinatorial rank is
\be
    \operatorname{cr}(\OGr_R(4,8),D)=\dim \Omega^6_{\log}(\OGr_R(4,8)\!\setminus\! D)=6 \,.
\ee
A basis of $\Omega^6_{\log}(\OGr_R(4,8)\!\setminus\! D)$ is given by the pullbacks under the isomorphism $\phi$ from \eqref{eq: iso with line arr} of the canonical forms of the four regions $R^{++},R^{+-},R^{-+},R^{--}$, together with the regions $R_1,R_2$ in Figure~\ref{fig: line arrangement}. The region $R_1$, for example, is bounded by the lines $\{ T, \hs S+T+U, \hs S+T-U\}$ and is identified projectively in Figure \ref{fig: line arrangement}. Similar statements hold for $R_2$.
The canonical forms of the additional two regions are
\begin{align}
    \Omega([R_1]) &= -\frac{S+T-U}{T(S+T+U)(S+T-U)}\:\dif S\wedge\dif T \,, \\[4pt]
    \Omega([R_2]) &=\frac{S+T-U}{S(S-T+U)(-S+T+U)}\:\dif S\wedge\dif T \,.
\end{align}
The linear independence of these forms can be verified by considering an arbitrary linear combination $\sum_{i=1}^{6}a_i \hs \Omega([R^{(i)}])$ and bringing all terms to a common denominator. Requiring this combination to vanish identically forces the resulting numerator to vanish as a polynomial in $S$, $T$, and $U$, which in turn implies $a_i=0$ for every $i$.

\section{Conclusions}
\label{sec:conclusions}
A remarkable success in the study of scattering amplitudes has been to reformulate $\mathcal{N}=4$ supersymmetric Yang--Mills theory in the framework of positive geometries. In this paper, we have taken the first steps in this direction in the context of cosmological correlators.

\vskip 4pt
We have discovered a geometric interpretation of Yang--Mills three- and four-point correlators at tree level within the cosmological Grassmannian. To this end, an important step has been to describe the latter in terms of Pfaffian coordinates, which provided a much more convenient embedding than the usual Plücker coordinates.
At four points, we have found both the reduced and the full color-ordered correlators. Interestingly, the latter is one of the first examples in physics of a positive geometry where the mixed Hodge theory approach was essential.\footnote{To our knowledge, the only other instance of a linear combination of positive geometries appears in the logarithm of loop-level scattering amplitudes in $\mathcal{N} = 4$ super Yang--Mills~\cite{Arkani-Hamed:2013kca, Arkani-Hamed:2021iya}. The related works, however, precede the formulation of the mixed Hodge theory approach. We thank Jaroslav Trnka for pointing this out to us.}

\vskip 4pt
There are several promising directions for future research:
\begin{itemize}
    \item The most immediate direction is to extend the construction to higher-point correlators. The recursive structure of canonical forms suggests the possibility of constructing higher-point geometries directly from lower-point ones. Developing such a geometric recursion could turn the present characterization into a practical method for computing correlators.

    \item A key step in our analysis was to allow for positive geometries to be linear combinations of cycles. This insight deserves further study, especially in connection with color relations and other linear identities among observables. This may clarify the physical significance of cycles that do not correspond to individual positive regions.

    \item A long-term goal is to extend the construction to loop-level correlators. It would be interesting to determine whether loop integrands admit a similar characterization in terms of canonical forms and whether new types of positive geometries are required.
\end{itemize}
It is very exciting to see that cosmological correlators also admit a positive-geometric description in a Grassmannian, and that Yang--Mills again finds a special place among all physical theories. We believe this is just the stepping stone in a more systematic understanding of geometric structures in the cosmological Grassmannian, and we look forward to its future developments.

\paragraph{Acknowledgements}
We thank Daniel Baumann, Bernd Sturmfels and Jaroslav Trnka for insightful discussions and comments on the draft. We also thank Daniel Baumann for mentioning this work during his talk at the ``Amplitudes 2026'' conference at Queen Mary University of London (QMUL). 

\vskip 4pt
The research of MA is funded by the European Union (ERC,  \raisebox{-2pt}{\includegraphics[height=0.9\baselineskip]{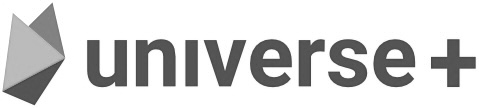}}, 101118787). FR is supported by the ERC (NOTIMEFORCOSMO, 101126304), by Scuola Normale, and by INFN (IS GSS-Pi). All authors received additional support from the European Union (ERC,  \raisebox{-2pt}{\includegraphics[height=0.9\baselineskip]{Figures/universe+_logo-typo-BW.jpg}}, 101118787).\footnote{\tiny Views and opinions expressed are however those of the author(s) only and do not necessarily reflect those of the European Union or the European Research Council Executive Agency. Neither the European Union nor the granting authority can be held responsible for them.}

\newpage
\appendix
\section{Grassmannian as Pure Spinors}\label{app: Spinors}
In this appendix, we explain how to describe the elements of the cosmological Grassmannian $\OGr(n,2n)$ with projective spinors that satisfy so-called purity constraints, rather than with $n\times 2n$ matrices $C$ and their minors. As we will see, the homogeneous coordinates of these pure spinors are Pfaffians of principal submatrices of $C$, which are, in a consistent sense, ``square roots'' of the minors of $C$. This alternative description will be much more economical as there are significantly fewer Pfaffians ($2^{n-1}$) than minors $\big(\binom{2n}{n}\big)$.

\vskip 4pt
We will first introduce the Clifford algebra in $\mathbb R^{n,n}$ using the conventional Gamma matrices $\Gamma^I$. We will then note that this algebra is equivalent to the algebra of $n$ fermionic oscillators, and show that spinors are simply arbitrary states in the corresponding Fock space, which may be expressed as linear combinations of multi-particle states.
Finally, we will define pure spinors and show how they describe elements of the cosmological Grassmannian. To make this construction accessible for both mathematicians and physicists, we will explicitly bridge the standard terminology used in both communities.

\subsection{Clifford Algebras}
We start by introducing the square Gamma matrices $\Gamma^I$ in $2n$ dimensions, with $I=\bar 1,\bar 2,\cdots,\bar n$, $1,2,\cdots,n$. These matrices satisfy the following relations:
\be 
    \label{equ:clifford}
    \{\Gamma^I,\Gamma^J\}=2\hs Q^{IJ} \hs \mathbf{1} \,, \quad\quad Q=\begin{pmatrix}
    0&1_{n\times n}\\
    1_{n\times n}&0
    \end{pmatrix},
\ee 
where $\mathbf{1}$ is the identity, and $\{A,B\}=AB+BA$ is the anticommutator.  The $\Gamma^I$ can be realized as explicit $2^n\times 2^n$ matrices. For instance, we can concretely define them in terms of the usual Pauli matrices $\sigma_3$ and $\sigma_\pm=(\sigma_1\pm i \sigma_2)/2$ as
\be 
    \label{equ:gamma-concrete}
    \begin{aligned}
    \Gamma^{\bar i}&=\sqrt{2}~\sigma_3\otimes \cdots\otimes \sigma_3\otimes \sigma_+\otimes 1_{2\times 2}\otimes\cdots \otimes 1_{2\times 2} \,,\\
    \Gamma^{ i}&=\sqrt{2}~\sigma_3\otimes \cdots\otimes \sigma_3\otimes \sigma_-\otimes 1_{2\times 2}\otimes\cdots \otimes 1_{2\times 2}\,,
\end{aligned}
\ee
where $i=1,2,\cdots,n$, and $\sigma_\pm$ are in the $i$-th place of the tensor product of $n$ matrices of size $2\times 2$.

\vskip 4pt
Alternatively, we can think of $\Gamma^I$ not as matrices, but as abstract generators equipped with a multiplication. The Clifford algebra in $2n$ dimensions, denoted by ${\rm Cl}(n,n)$, is then defined as the (real) algebra generated by $\Gamma^I$, that is, the algebra comprising the identity $\mathbf{1}$, the generating elements~$\Gamma^I$, and any linear combination of products of the~$\Gamma^I$, such that the relation \eqref{equ:clifford} holds with the metric $Q^{IJ}$.
Formally, the Clifford algebra as defined above is the quotient of the tensor algebra \cite[Chapter 1]{spingeo}:
\be\label{eq: Cliff def}
    {\rm Cl}(n,n) \equiv T^\bullet V / I\,, \quad\quad I\equiv \langle \Gamma^I\otimes \Gamma^J +\Gamma^J\otimes \Gamma^I -2\hs Q^{IJ}\hs \mathbf{1}\rangle \,.
\ee
Here, $V$ is a $2n$-dimensional (real) vector space spanned by the generators $\Gamma^I$, and $T^\bullet V$ is the tensor algebra $T^\bullet V=\oplus_{k=0}^\infty V^{\otimes k}$. Indeed, the multiplication is realized in the tensor algebra as a tensor product, and the relation \eqref{equ:clifford} is enforced in the quotient by the ideal $I$ generated by $\Gamma^I\otimes \Gamma^J +\Gamma^J\otimes \Gamma^I -2\hs Q^{IJ}\hs \mathbf{1}$. 
Note that the concrete choice of $\Gamma^I$-matrices is a representation of the Clifford algebra. Since there is a unique irreducible representation of ${\rm Cl}(n,n)$ as matrices of dimension $2^n\times 2^n$~\cite{Freedman:2012zz}, the choice of $\Gamma^I$-matrices in \eqref{equ:gamma-concrete} with the usual matrix multiplication is essentially unique up to conjugation, $\Gamma^I\mapsto M \cdot \Gamma^I \cdot M^{-1}$.

\vskip 4pt
Note that any symmetric product of $\Gamma^I$-matrices can be written in terms of a product of fewer $\Gamma^I$-matrices due to the relation \eqref{equ:clifford}, hence the only independent products in the Clifford algebra are the antisymmetric products of $r$ $\Gamma^I$-matrices, with $0\leq r\leq 2n$. We label them as
\be 
\Gamma^{I_1\cdots I_r}\equiv \Gamma^{[I_1}\Gamma^{I_2}\cdots \Gamma^{I_r]}\equiv\frac{1}{r!}\sum_{\sigma\in S_r}\text{sign}(\sigma) \hs \Gamma^{I_{\sigma(1)}}\Gamma^{I_{\sigma(2)}}\cdots \Gamma^{I_{\sigma(r)}}\,.
\ee 
By convention, this is the identity $\mathbf{1}$ for $r=0$.
Altogether, these $\sum_{r=0}^{2n}\binom{2n}{r}=2^{2n}$ antisymmetric products form a basis of the Clifford algebra ${\rm Cl}(n,n)$.
Realizing the $\Gamma^I$-matrices as $2^n\times 2^n$ matrices exhibits the isomorphism of the Clifford algebra with the space of $2^n\times 2^n$ matrices. Indeed, given any $2^n\times 2^n$ matrix $M$, it can be written as a linear combination of the matrices $\{\mathbf{1},\Gamma^I,\Gamma^{I_1I_2},\cdots,\Gamma^{I_1\cdots I_{2n}}\}$:
\be  \label{equ:expansion-gammas}
    M=m_0\hs  1_{2^n\times 2^n}+m_I \Gamma^I+\frac{1}{2!}m_{I_1I_2}\Gamma^{I_1I_2}+\cdots+\frac{1}{(2n)!}m_{I_1\cdots I_{2n}}\Gamma^{I_1\cdots I_{2n}} \,.
\ee 
The $\Gamma^I$-matrices and all their antisymmetric products satisfy an orthogonality property~\cite{Freedman:2012zz}:
\be 
    \text{Tr}(\Gamma^{I_1\cdots I_r} \cdot \Gamma_{J_kJ_{k-1}\cdots J_1})=\delta_{rk}\hs 2^n\hs r! \hs \delta_{[J_1}^{~I_1}\cdots \delta_{J_r]}^{~I_r}\,, \qquad \Gamma_{J_k\cdots J_1} \equiv Q_{J_kK_k} \cdots Q_{J_1K_1}\Gamma^{K_k\cdots K_1}\,.
\ee 
This can be checked explicitly in the representation \eqref{equ:gamma-concrete}. Hence, each coefficient in this expansion is given by
\be 
    \label{equ:matrix-coeff}
    m_{I_1\cdots I_r}=\frac{1}{2^n}\text{Tr}(M \cdot \Gamma_{I_r\cdots I_1}) \,.
\ee
Note that the $\Gamma^I$-matrices naturally represent linear maps from a $2^n$-dimensional (real) vector space $S$ to itself. We can interpret $S$ as the space of (Dirac) spinors, where a spinor is just a $2^n$-dimensional vector $\Psi$ and the $\Gamma^I$-matrices act by multiplication.

\vskip 4pt
Moreover, if we only consider the antisymmetric product of two $\Gamma^I$-matrices, we obtain a space of matrices closed under the usual commutator bracket. Indeed, if we normalize $\Sigma^{IJ}= \Gamma^{IJ}/2$, we obtain the standard generators of the Lie algebra $\frak{so}(n,n) \cong \frak{spin}(n,n)$, as they satisfy the commutation relations
\be [\Sigma^{IJ},\Sigma^{KL}]=Q^{JK}\Sigma^{IL}-Q^{IK}\Sigma^{JL}-Q^{JL}\Sigma^{IK}+Q^{IL}\Sigma^{JK}\,.
\ee 
In this setup, considering finite products of exponential series of elements of $\frak{spin}(n,n)$ yields the spin group ${\rm Spin}(n,n)$. Note that this exhibits ${\rm Spin}(n,n)$ as a subgroup of the Clifford algebra. We can then consider the space of Dirac spinors as a representation of the spin group. As we will review, it can be decomposed into two irreducible representations: the left and right (Weyl) spinors. Let us now turn to an alternative realization of the Clifford algebra as the algebra of creation and annihilation operators. This will be especially useful for expressing spinors in a convenient basis of multi-particle states.

\subsection{Fermionic Oscillators}
The Clifford algebra~\eqref{equ:clifford} is the same as the algebra of $n$ fermionic oscillators, generated by the \textit{creation} and \textit{annihilation} operators $\big(a^i\big)^\dag $ and $a^i$, with $i=1,2,\cdots,n$. These operators satisfy the following relations:
\be \label{equ:algebra-fermion}
\{a^i,\big(a^j\big)^\dag\}=\delta^{ij}\,,\quad\quad\{a^i,a^j\}=\{\big(a^i\big)^\dag, \big(a^j\big)^\dag \}=0\,.
\ee 
Indeed, after identifying the matrices $\Gamma^I=(\Gamma^{\bar i},\Gamma^{i})$ with 
\be 
\label{equ:aGammaident}
\Gamma^{\bar i}\leftrightarrow \sqrt{2}\hs a^i \,,\qquad \Gamma^{ i}\leftrightarrow \sqrt{2}\hs \big(a^i\big)^\dag \,,
\ee 
the algebra defined by the relations \eqref{equ:algebra-fermion} coincides with the Clifford algebra given by \eqref{equ:clifford}.

\vskip 4pt
Thus, the creation and annihilation operators act naturally on the space of spinors $S$. We can identify the \textit{vacuum state} $\Psi^0$ as the non-trivial vector annihilated by all the $a^i$, i.e.~$a^i\Psi^0=0$ for all $i$.\footnote{This vacuum state always exists by construction. Take a non-trivial spinor $v$. If it is not annihilated by some~$a^i$, we can construct a new non-trivial spinor $v'=a^i v$ that satisfies $a^i v'=0$ due to the anticommutation relations $\{a^i,a^i\}=2 (a^i)^2=0$. Repeating this reasoning for each of the $n$ annihilation operators $a^i$, one can always construct a non-trivial vacuum state annihilated by $(a^i)$ for all $i$.} We then construct all the ``multi-particle'' states by an iterative application of the creation operators as\footnote{These multi-particle states are straightforward in the explicit matrix realization \eqref{equ:gamma-concrete} of the $\Gamma^I$-matrices. Indeed, we can define $|0\rangle\equiv (1,0)$ and $|1\rangle\equiv (0,1)$ as the states with $0$ and $1$ fermion, respectively. Then, the vacuum is simply $\Psi^0=|0\rangle\otimes \cdots\otimes |0\rangle$, and the multi-particle states are explicitly given by
\[ 
\Psi^{i_1\cdots i_k}=(a^{i_1})^\dag (a^{i_2})^\dag \cdots (a^{i_k})^\dag \Psi^0=|0\rangle\otimes \cdots\otimes  |0\rangle\otimes \underbrace{|1\rangle}_{i_1}\otimes\cdots |0\rangle\otimes\cdots \otimes \underbrace{|1\rangle}_{i_k}\otimes\cdots\otimes |0\rangle\,,
\]
where we assume that $i_1<i_2<\cdots <i_k$, and use that $|1\rangle=\sigma_-|0\rangle$. The resulting state is the tensor product of $n$ states that are $|1\rangle$ for the fermionic oscillators in the subset $\{i_1,i_2,\cdots ,i_k\}$, and $|0\rangle$ for the remaining oscillators.}
\be 
\Psi^{i_1\cdots i_k}= \big(a^{i_1i_2\cdots i_k} \big)^\dag \Psi^0 \,,\quad\quad \big(a^{i_1i_2\cdots i_k}\big)^\dag\equiv \big(a^{i_1}\big)^\dag \big(a^{i_2}\big)^\dag \cdots \big(a^{i_k}\big)^\dag\,,
\ee 
which is antisymmetric in $i_1\cdots i_k$. These multi-particle states form a particular basis for the vector space of spinors $S$. We denote the number operators of each fermionic particle $i=1,2,\cdots,n$ by $N_i\equiv a_i^\dag a_i$. In the basis $\Psi^{i_1\cdots i_k}$, the matrices corresponding to $N_i$ are simultaneously diagonalized. Moreover, since \eqref{equ:algebra-fermion} implies $N_i(1-N_i)=0$, the eigenvalues of each number operator $N_i$ can only be $1$ or $0$, which corresponds to having one or zero particles of type $i$. Concretely, the state $\Psi^{i_1\cdots i_k}$ is an eigenvector of $N_i$ with eigenvalue $1$ or $0$ depending on whether $i \in \{i_1,\cdots,i_k\}$ or~not, respectively.

\vskip 4pt
We can therefore write any spinor as the following linear combination of multi-particle states
\begin{align}\label{equ:spinor-psi}
    \Psi&= \left(p_{0}+ p_i \big(a^i\big)^\dag+\frac{1}{2!}\hs p_{i_1i_2} \big(a^{i_1i_2}\big)^\dag+\cdots+\frac{1}{n!}\hs p_{i_1i_2\cdots i_n} \big(a^{i_1i_2\cdots i_n}\big)^\dag\right)\Psi^0 \nonumber\\
    &=p_{0}\Psi^0+ p_i \Psi^i+\frac{1}{2!}\hs p_{i_1i_2}\Psi^{i_1i_2}+\cdots+\frac{1}{n!}\hs p_{i_1i_2\cdots i_n}\Psi^{i_1i_2\cdots i_n}\,.
\end{align}
Note that there are $\displaystyle\sum_{k=0}^n\binom{n}{k}=2^n$ coefficients~$p_{i_1\cdots i_r}$, compatibly with the fact that the spinor $\Psi$ has $2^n$ components. This also justifies our assumption that $\Psi^0$ is the unique vacuum.

\vskip 4pt
\paragraph{Chirality} The chirality operator of the Clifford algebra $\Gamma_*$ has a nice representation in terms of the total number operator $N= N_1+\cdots +N_n$:
\be 
    \Gamma_* \equiv \Gamma^{\bar 11\bar 22\cdots \bar n n} = \Gamma^{[\bar 1}\Gamma^1\Gamma^{\bar 2}\Gamma^2\cdots \Gamma^{\bar n} \Gamma^{n]} =(-1)^N \,,
\ee 
where we used $\Gamma^{[\bar i}\Gamma^{i]}=(a_i a_i^\dag -a_i^\dag a_i) = (1-2 N_i) = (-1)^{N_i}$. The chirality operator has eigenvalues $-1$ and $+1$ and splits the space of spinors into its two eigenspaces, the left and right Weyl spinors. These are defined by
\be 
    \Gamma_*\Psi_L=-\Psi_L \,,\quad \text{and}\quad\Gamma_* \Psi_R=+\Psi_R\,.
\ee 
Since $\Gamma_*=(-1)^N$, the space of right (left) spinors consists of those multi-particle states that have an even (odd) number of fermions, respectively. Thus, the multi-particle states provide a basis for the space of spinors where chirality is manifest. Indeed, a generic Dirac spinor~\eqref{equ:spinor-psi} can be split into its left and right parts as $\Psi=\Psi_L+\Psi_R$, with an odd and even number of fermions respectively, and
\be
\begin{array}{r@{\:}c@{\:}c@{\:+\:}c@{\:+\:}c@{\:+\:}c}
\Psi_R & = & p_{0}\Psi^0 & \displaystyle \frac{1}{2!}\hs p_{i_1i_2}\Psi^{i_1i_2} & \displaystyle \frac{1}{4!}\hs
p_{i_1i_2i_3i_4}\Psi^{i_1i_2i_3i_4} & \cdots \,, \\[10pt]
\Psi_L & = & p_i\Psi^i & \displaystyle \frac{1}{3!}\hs
p_{i_1i_2i_3}\Psi^{i_1i_2i_3} & \displaystyle \frac{1}{5!}\hs p_{i_1i_2i_3i_4i_5}\Psi^{i_1i_2i_3i_4i_5} & \cdots \,.
\end{array}
\label{equ:psiLR}
\ee
Here, the sums in $\Psi_R$ and $\Psi_L$ range over all states with an even and an odd number of fermions, respectively, up to fermion number $n-1$ or $n$, depending on the parity of $n$.

\vskip 4pt
Since the chirality operator $\Gamma_*$ commutes with all the $\mathfrak{spin}(n,n)$ generators $\Sigma_{IJ}$, their action on a Weyl spinor must give a Weyl spinor with the same chirality. Hence, when exponentiating these generators, chirality is preserved by the action of the spin group. Thus, the space of Dirac spinors is reduced to the direct sum of the spaces of Weyl-right and Weyl-left spinors as $S=S_R\oplus S_L$, which are in turn irreducible representations of the spin group of dimension $2^{n-1}$. These are sometimes called the \emph{half-spin representations}.

\subsection{Pure Spinors}
In this subsection, we will explain the correspondence between (projective) pure spinors and the cosmological Grassmannian~$\OGr(n,2n)$, i.e.~the set of $n$-planes in $\RR^{2n}$ that are null with respect to $Q$. Remarkably, this yields a minimal embedding of $\OGr(n,2n)$ which is also natural from a mathematical point of view.

\vskip 4pt
Recall that, given a spinor $\Psi$, its Majorana conjugate $\bar\Psi$ is given by the condition that $\bar \Psi \Psi$ is invariant under the action of ${\rm Spin}(n,n)$. More explicitly, we define it as~\cite{VanProeyen:1999ni} 
\be \label{eq: psibar}
\bar\Psi \equiv \Psi^T \hs B\,, \quad\text{where}\quad B\equiv (\Gamma^1+\Gamma^{\bar 1})\cdots (\Gamma^n+\Gamma^{\bar n})\,.
\ee 
The Majorana conjugate is such that every matrix $B \hs \Gamma^{I_1...I_r}$ is either symmetric or antisymmetric~\cite{VanProeyen:1999ni}.
We say that $\Psi$ is a \emph{pure spinor} if the expansion \eqref{equ:expansion-gammas} of its ``square'' $\Psi\bar\Psi$ in terms of the Gamma matrices only contains the antisymmetric product of exactly $n$ Gamma matrices, i.e.
\be \label{equ:purespinor-M}
\Psi  \bar\Psi= \frac{1}{n!}\hs\Gamma^{I_1\cdots I_n}\hs m_{I_1\cdots I_n}\,.
\ee 
Here, $\Psi$ and $\bar\Psi$ are column and row vectors, respectively, therefore $\Psi\bar\Psi$ is a matrix, while $\bar\Psi\Psi$ is a number.
Using \eqref{equ:matrix-coeff}, this is equivalent to the \emph{purity constraints}
\be \label{equ:purity-constraints}
    \bar\Psi \cdot \Gamma_{I_1\cdots I_r} \cdot \Psi=0\quad\quad\forall \hs r\neq n\,.
\ee 
These are a set of quadratic constraints on the components~$p_{i_1\cdots i_r}$ of the spinors $\Psi$. For example, for $n=4$, the only non-trivial constraint is $\bar\Psi\Psi=0$, which is explicitly given by
\be \label{equ:purity-constraint-n=4}
    p_{12} \hs p_{34} - p_{13} \hs p_{24} + p_{14} \hs p_{23}- p \hs p_{1234}=0\,.
\ee
The purity condition~\eqref{equ:purespinor-M} allows us to identify the coefficients $m_{I_1\cdots I_n}$ with the minors of a certain $n \times 2n$ matrix $C$ that satisfies $C\cdot Q \cdot C^T=0$ as
\be \label{equ:minor-C}
m_{I_1\cdots I_n}=(I_1\cdots I_n)=\epsilon^{a_1\cdots a_n}C_{a_1I_1}\cdots C_{a_n I_n}\,.
\ee 
To prove this, one needs to check that they satisfy the Plücker and the orthogonality relations~\cite{elmaazouz2025positive}. Indeed, note that $\Gamma^I\Gamma_{I_1\cdots I_n}\Gamma_I=0$ and the purity condition~\eqref{equ:purespinor-M} imply $\Gamma^I \Psi\bar\Psi \Gamma_I=0$. 
Multiplying this on the right by $\Gamma_{I_1\cdots I_{n-1}}\Psi$ and using $\Gamma_I\Gamma_{I_1\cdots I_{n-1}}=\Gamma_{II_1\cdots I_{n-1}}+(n-1)Q_{I[I_1}\Gamma_{I_2\cdots I_{n-1}]}$, together with the purity constraints~\eqref{equ:purity-constraints}, we obtain
\be \label{equ:intermediate}
    \Gamma^I \Psi\bar\Psi \Gamma_{II_1\cdots I_{n-1}}\Psi=0 \implies m_{II_1\cdots I_{n-1}}\Gamma^I\Psi=0\,.
\ee 
Further multiplying this on the right by $\bar\Psi$, we get $m_{II_1\cdots I_{n-1}}m_{J_1\cdots J_n}\Gamma^I\Gamma^{J_1\cdots J_n}=0$.
Expanding the resulting matrix in the basis~\eqref{equ:expansion-gammas}, and using~\eqref{equ:matrix-coeff} to compute the coefficients of the Gamma matrices $\Gamma^{K_1\cdots K_{n+1}}$ with $n+1$ indices, we obtain all the Plücker relations:
\be 
    m_{I_1\cdots I_{n-1}[K_1}m_{K_2\cdots K_{n+1}]}=0\,.
\ee 
Moreover, Equation~\eqref{equ:intermediate} implies all the orthogonality relations as
\be 
    m_{II_1\cdots I_{n-1}}m_{JJ_1\cdots J_{n-1}}\{\Gamma^I,\Gamma^J\}\Psi=0 \implies m_{II_1\cdots I_{n-1}}m_{JJ_1\cdots J_{n-1}} Q^{IJ}=0\,,
\ee
where we have further multiplied on the left by $m_{J J_1 \cdots J_{n-1}} \Gamma^J$ and symmetrized. Since they satisfy the required relations, the coefficients $m_{I_1\cdots I_n}$ can therefore be identified with the minors $(I_1\cdots I_n)$ of a matrix $C$ in $\OGr(n,2n)$, as claimed. In this way, the pure spinor is a ``square root'' of the minors:
\be \label{equ:purespinor}
    \Psi\bar\Psi=\frac{1}{n!}\hs\Gamma^{I_1\cdots I_n}\hs (I_1\cdots I_n) \implies (I_1\cdots I_n)=\frac{1}{2^{n}} \bar\Psi \cdot \Gamma_{I_n\cdots I_1} \cdot \Psi\,.
\ee 
This relation allows us to express any minor $(I_1\cdots I_n)$ as a homogeneous, quadratic polynomial of the spinor components. For $n=4$, for instance, we can compute the following minor:
\be \label{equ:minor-S}
    S=(\bar1\bar212)=\frac{1}{2^4}\bar\Psi \cdot \Gamma_{21\bar2\bar1} \cdot \Psi=-\frac{1}{2^4}\bar\Psi (-1)^{N_1+N_2}\Psi=\frac{1}{2}(p_{12}\hs p_{34}-p \hs p_{1234}+p_{13}\hs p_{24}-p_{14}\hs p_{23})\,,
\ee
which may be written as
$S= p_{12}\hs p_{34}-p\hs  p_{1234}=p_{13} \hs p_{24} - p_{14} \hs p_{23}$ due to the purity constraint~\eqref{equ:purity-constraint-n=4}.

\vskip 4pt
The purity constraints \eqref{equ:purity-constraints} are invariant under rescalings of $\Psi$. We will therefore always consider \textit{projective} pure spinors from now on, so that $\Psi \sim \rho \Psi$ for $\rho \in \RR^*$.
This rescaling maps the minors $(I_1 \cdots I_n)$ to $\rho^2 (I_1 \cdots I_n)$, which determine the same element of $\ogr(n,2n)$. For any projective pure spinor, Equation~\eqref{equ:purespinor} allows us to construct an $n$-plane in $2n$ dimensions, spanned by the rows of the matrix $C$, that is null with respect to the metric $Q$.\footnote{This is called a $Q$-isotropic subspace in the mathematics literature.}
As we show below, we can also construct a projective pure spinor out of any such plane, hence this turns out to be a one-to-one correspondence between the space of projective pure spinors and the orthogonal Grassmannian $\OGr(n,2n)$~\cite{Berkovits:2004bw}.

\vskip 4pt
Alternatively, we can associate a plane $S_\Psi$ in $2n$ dimensions to any spinor $\Psi$ as the space of vectors $v$ in $2n$ dimensions, with components $v_I$, such that $\slashed{v}\Psi=0$, where $\slashed{v}\equiv v_I\Gamma^I$. This plane $S_\Psi$ must be null:
\be \label{equ:vw-null}
2\hs  v_I Q^{IJ}  w_J\hs \Psi=\{\slashed{v},\slashed{w}\}\Psi=0\implies v_I Q^{IJ}w_J=0\,,
\ee 
for any $v,w\in S_\Psi$. A spinor can equivalently be defined to be pure if its associated null plane $S_\Psi$ has dimension $n$, which is precisely the same null $n$-plane as the one characterized by the minors in Equation~\eqref{equ:purespinor}.\footnote{We show the equivalence of both definitions. First, let $\Psi$ be pure in the sense of \eqref{equ:purespinor-M}, and $C=C_{aI}$~be the matrix entering the minors in~\eqref{equ:purespinor}. Using \eqref{equ:minor-C} and $\{\slashed C_a,\slashed C_b\}=0$, which follows from~$C \cdot Q \cdot C^T=0$, we have
\[
C_{aI}\Gamma^I\Psi\bar\Psi
=\frac{1}{n!}\epsilon^{a_1\cdots a_n}\slashed C_a\slashed C_{a_1}\cdots\slashed C_{a_n}=0\, .
\]
Every row of $C$ then lies in $S_\Psi$, so $S_\Psi\supset\mathrm{rowspan}(C)$. Since $S_\Psi$ is null, $\dim S_\Psi\leq n=\dim\mathrm{rowspan}(C)$, and hence $S_\Psi=\mathrm{rowspan}(C)$ is $n$-dimensional. Conversely, assume $\dim S_\Psi=n$ and take any $\Gamma^{I_1 \ldots I_r}$, with $r<n$. By dimensionality, there exists $w\in S_\Psi$ such that $w^{I_j}=0$ for $j=1,\ldots,r$. Choosing $u\in\RR^{2n}$ such that $u_IQ^{IJ}w_J=1/2$, and using $\{\slashed w,\Gamma^{I_j}\}=0$, $\{\slashed w,\slashed u\}=1$ and $\slashed w\Psi=0$ (which further implies $\bar \Psi\slashed w=0$ by the properties of $B$ in \eqref{eq: psibar} under transposition), we find
\[
\bar\Psi \cdot \Gamma^{I_1\cdots I_r} \cdot \Psi
= \bar\Psi \cdot \Gamma^{I_1\cdots I_r}(\slashed w\slashed u+\slashed u\slashed w) \cdot \Psi
=(-1)^r\bar\Psi \cdot \slashed w\Gamma^{I_1\cdots I_r}\slashed u \cdot \Psi=0 \,.
\]
To prove~\eqref{equ:purity-constraints} also for $r>n$, we can use $\Gamma^J\Gamma^{I_2\cdots I_r}
+
(-1)^{r}\Gamma^{I_2\cdots I_r}\Gamma^J
=
2\Gamma^{J I_2\cdots I_r}$ to show that $w_{I_1}\bar\Psi\cdot\Gamma^{I_1\cdots I_r}\cdot\Psi=0$ vanishes for any $w_{I_1}\in S_\Psi$.
The product $\bar\Psi\cdot\Gamma^{I_1\cdots I_r}\cdot\Psi$ therefore descends to an $r$-form (i.e.~an antisymmetric tensor) on the orthogonal complement $(S_\Psi)^\perp = Q \cdot S_\Psi$, which has dimension $n$. Since $r>n$, such an $r$-form must vanish.}

\vskip 4pt
It can be proven that a pure spinor must be a Weyl spinor.\footnote{Indeed, using $(-1)^n\Gamma_*\Gamma^{I_1\cdots I_n}\Gamma_* =\Gamma^{I_1\cdots I_n}$ together with the properties of the Gamma matrices under transposition~\cite{VanProeyen:1999ni}, it is straightforward to show that the map $\Psi\mapsto \Gamma_*\Psi$ preserves the matrix $\Psi\bar\Psi=\Gamma^{I_1\cdots I_n}(I_1\cdots I_n)/n!$ for a pure spinor. Thus, since $(\Gamma_*\Psi)(\overline{\Gamma_*\Psi})=\Psi\bar\Psi$, this implies that $\Gamma_* \Psi=\pm \Psi$, i.e.~that the pure spinor $\Psi$ must be a Weyl spinor.}
Thus, there are two connected components in the space of (projective) pure spinors: the Weyl-right and Weyl-left spinors. As we show below, these correspond in turn to the two branches of the orthogonal Grassmannian $\OGr(n,2n)$, which are isomorphic to each other. Our presentation will focus on the right branch, which we denote as $\ogr_R(n,2n)$.

\vskip 4pt
For concreteness, let us consider an example of a pure spinor. By definition, the vacuum state $\Psi^0$ is annihilated by all the operators $a^i$. Thus, its associated plane $S_{\Psi^0}$ is $n$-dimensional as it is spanned by the $n$ rows of the matrix $C_0=(1_{n\times n},0_{n\times n})$. Indeed, the operators $(C_0)_{aI} \Gamma^I$ correspond to the $n$ annihilation operators $a^i$ through \eqref{equ:aGammaident}. The vacuum state is therefore a Weyl-right pure spinor whose corresponding element in $\OGr_R(n,2n)$ is precisely $S_{\Psi^0}$. 

\vskip 4pt
We can use the action of the spin group on the vacuum state to construct more pure spinors of the same chirality.
First, note that applying any transformation $g$ of the group Spin$(n,n)$ to any pure spinor $\Psi$ will yield another pure spinor $g\Psi$.
The reason is that, if we map $\Psi\mapsto g\Psi$ in Equation~\eqref{equ:purespinor}, the resulting right-hand side is also a combination of Gamma matrices with exactly $n$ indices because the matrices $\Gamma^{I_1\cdots I_n}$ transform as an antisymmetric tensor with respect to the spin group~\cite{Freedman:2012zz}.
Moreover, if the pure spinor $\Psi$ is Weyl-right, the transformed pure spinor $g\Psi$ is also Weyl-right because chirality is preserved as well. 
In particular, since the vacuum is a Weyl-right pure spinor, $g \Psi^0$ must also be one.

\vskip 4pt
We now explicitly determine the orbit of the vacuum under the action of the spin group.
On a dense open subset, elements of $\operatorname{Spin}(n,n)$ admit the decomposition~\cite{Balian:1969tb, Mirjafarlou:2023tbi}
\be
     g =
    \exp\bigg(\frac{1}{2}c_{ij}(a^i)^\dag(a^j)^\dag\bigg)
    \exp\bigg(d_{ij}[(a^{i})^\dag, a^{j}]\bigg)
    \exp\bigg(\frac{1}{2}e_{ij}a^ia^j\bigg) \,.
\ee
The last two factors preserve the projective spinor $\Psi^0$ and therefore belong to its stabilizer
\be
    P=\big\{h\in\operatorname{Spin}(n,n):h\Psi^0\sim\Psi^0\big\} \,.
\ee
Consequently, the orbit of $\Psi^0$ is given by
\be
    \label{equ:Psi-gPsi0}
    \Psi(c_{ij})=g(c_{ij})\Psi^0\,,\quad\text{where}\quad g(c_{ij})\equiv\exp\bigg(\frac{1}{2} c_{ij}\big(a^i\big)^\dag\big(a^j\big)^\dag\bigg)\,.
\ee 
This is isomorphic to the dense open cell of $\operatorname{Spin}(n,n)/P$ parameterized by the elements $g(c_{ij})$. 
Moreover, note that $\Psi(c_{ij})$ is annihilated by the $n$ operators $g a^ig^{-1}=a^i-c_{ij} \big(a^j\big)^\dag$.
Thus, it is a pure spinor associated with the plane spanned by the $n$ rows of
\be \label{equ:matrix-C-appendix}
    C=\begin{pmatrix}
      1_{n\times n},& C_n  
    \end{pmatrix},
\ee 
with $(C_n)_{ij}=-c_{ij}$, which is an element of $\OGr_R(n,2n)$. Indeed, it is straightforward to check that $C_{aI}\Gamma^I \Psi=0$.
Furthermore, note that any element in (the affine chart of) $\OGr_R(n,2n)$ is given by a generic matrix \eqref{equ:matrix-C-appendix}, and it is therefore associated with a pure spinor that can be obtained from a spin transformation applied to the vacuum.\footnote{Using the fermionic oscillator analogy, the transformation $g(c_{ij})$ can be interpreted as a fermionic Bogoliubov transformation that maps the vacuum state $\Psi^0$ to another state $\Psi$ annihilated by the $n$ operators $a^i-c_{ij}\big(a^j\big)^\dag$.
The latter may be seen as the transformed annihilation operators.}  
Since the affine chart is dense, this association of elements extends to the closures of each set, yielding the isomorphisms
\be \label{eq:closures}
    \ogr_R(n,2n) \cong {\rm Spin}(n,n)/P \cong \overline{\{g(c_{ij}) \Psi^0 \}}\,.
\ee
Equivalently, we can associate any matrix $C_{aI}$ in $\ogr_R(n,2n)$, which may not belong to this affine chart, to the unique projective spinor annihilated by the $n$ operators $C_{aI}\Gamma^I$. The latter is a pure spinor by the definition given below Equation~\eqref{equ:vw-null}. This yields an isomorphism between $\ogr_R(n,2n)$ and the space of Weyl-right projective pure spinors, together with the inverse map~\eqref{equ:purespinor}. As a consequence, we find that the latter coincides with the closure of the orbit $\overline {\{ g(c_{ij}) \Psi^0 \}}$, which is in turn the closed set $\{ g \Psi^0 : g \in \mathrm{Spin}(n,n) \}$.

\vskip 4pt
We have just explicitly exhibited a very classical result in the theory of flag varieties for the case of $\OGr_R(n,2n)$. That is, given a flag variety $G/P$, we can embed it into the projectivization $\PP(W)$ of an irreducible representation $W$ of $G$. Moreover, there exists a projective vector $w_0$ in $\PP(W)$ such that $P$ is the stabilizer of $w_0$, i.e.~the subgroup of $G$ that leaves $w_0$ invariant, and $G/P$ is the closure of the $G$-orbit of $w_0$ in $\PP(W)$.  In our case, $G={\rm Spin}(n,n)$, $W=S_R$ is the space of right Weyl spinors, $w_0$ is the (projective) spinor corresponding to the vacuum state $\Psi^0$, and $P$ is the stabilizer of $\Psi^0$, so that $\OGr_R(n,2n)\simeq \mathrm{Spin}(n,n)/P$. In particular, we have shown that $\OGr_R(n,2n)$ corresponds to the projective pure spinors in $\PP(S_R)$. As the corresponding representation is (co)minuscule, this yields a minimal homogeneous embedding; see \cite{LandsbergManivel2003} for a detailed discussion. 

\vskip 4pt
We now want to express the spinor $\Psi=g\Psi^0$ explicitly as a right Weyl spinor in the basis of Equation~\eqref{equ:psiLR}. Expanding the exponential in Equation~\eqref{equ:Psi-gPsi0}, we find
\be\label{equ:Psi-p-cij}
    \Psi
    =\sum_{r=0}^{\floor{n/2}} \frac{1}{(2r)!} \hs p_{i_1i_2\cdots i_{2r}} \big(a^{i_1\cdots i_{2r}}\big)^\dag \Psi^0 \,,\quad \text{with} \quad p_{i_1\cdots i_{2r}}= \frac{(2r)!}{2^r \hs r!} \hs c_{[i_1i_2}\cdots c_{i_{2r-1}i_{2r}]}\,.
\ee
From Equation~\eqref{eq: Pfaffian}, the coefficient $p_{i_1\cdots i_{2r}}$ is precisely the Pfaffian of the principal submatrix indexed by the rows and columns $i_1,i_2,\ldots,i_{2r}$ of the $n\times n$ skew-symmetric matrix $-(C_n)_{ij}=c_{ij}$.
In fact, the Pfaffian relations~\eqref{eq: pfaffian rels} correspond precisely to the purity constraints \eqref{equ:purity-constraints} for the spinor $\Psi$ in \eqref{equ:Psi-p-cij}.
In the affine chart $p=1$, they fix the components $p_{i_1\cdots i_{2r}}$ in terms of $p_{ij}=c_{ij}$.
Although the Pfaffian parameterization \eqref{eq:spinorembed} applies when $p=1$, which corresponds to the affine chart \eqref{equ:matrix-C-appendix} of $\OGr_R(n,2n)$, projective pure spinors $\Psi=(p:p_{ij}:p_{ijkl}:\cdots)$ are precisely the closure of the orbit of $\Psi^0$ under the spin group, hence they describe all elements of $\OGr_R(n,2n)$ as shown in \eqref{eq:closures}. This shows that the Pfaffian parameterization indeed extends to an embedding of the orthogonal Grassmannian, and can naturally be lifted to its complexification.

\vskip 4pt
For $n=4$, for example, we have
\be 
    \Psi=\left(1+\frac{1}{2!}c_{ij} \big(a^{ij}\big)^\dag
    +(c_{12}c_{34}-c_{13}c_{24}+c_{14}c_{23}) \big(a^{1234}\big)^\dag\right)\Psi^0\,,
\ee 
where we explicitly summed over permutations in the last term. Thus, we may identify the pure spinor in the basis of multi-particle states $\{\Psi^{i_1\cdots i_{2r}}\}$ as
\be 
\Psi=(p:p_{12}:p_{13}:p_{14}:p_{23}:p_{24}:p_{34}:p_{1234})\,,
\ee 
where $p=1$, $p_{ij}=c_{ij}$ and $p_{1234}=c_{12} \hs c_{34}-c_{13} \hs c_{24}+c_{14} \hs c_{23}$. 
The purity constraint~\eqref{equ:purity-constraint-n=4} fixes $p_{1234}$ in the affine chart $p=1$ in terms of the other components as $p_{1234}=c_{12} \hs c_{34}-c_{13} \hs c_{24}+c_{14} \hs c_{23}$, which is the $4 \times 4$ Pfaffian of $-C_4$.

\newpage
\phantomsection
\addcontentsline{toc}{section}{References}
\bibliographystyle{utphys}
{\linespread{1.075}
	\bibliography{ref.bib}
}

\end{document}